%% file: main.tex
\documentclass[11pt]{article}
\usepackage[margin=1in]{geometry} 

\usepackage{graphicx} % Required for inserting images
\usepackage{algorithm}
\usepackage{algpseudocode}
\usepackage{authblk}
\usepackage{amsmath}
\usepackage{amsthm}
\usepackage{titlesec}
\usepackage{booktabs}
\usepackage{braket}

\title{An Information-Minimal Geometry for Qubit-Efficient Optimization}

\author[1]{Gordon Ma\thanks{gordonma@nus.edu.sg}}
\author[1,2,3,4]{Dimitris G. Angelakis \thanks{dimitris.angelakis@gmail.com}}

\affil[1]{Centre for Quantum Technologies, National University of Singapore, Singapore}
\affil[2]{Institute for Quantum Computing and Quantum Technologies, NCSR Demokritos, Greece}
\affil[3]{School of Electronics and Computer Science, University of Southampton, Southampton, UK}
\affil[4]{AngelQ Quantum Computing, Singapore}
\date{}

\input{preamble}

\begin{document}
\maketitle
\vspace{-3.3em}  % Adjust this value as needed
\input{sections/00-abstract}

\vspace{-1.5em}  % Adjust this value as needed
\input{sections/01-intro-restructure}
\input{sections/02-approaches-rev}

\input{sections/03-methods}
\input{sections/04-experiments}
\input{sections/09-discussion-outlook-epilogue}

\paragraph{Acknowledgements.}
The authors thanks Daniel Leykman, Vasileios Tsiaras and Bhuvanesh Sundar for fruitful discussions. We thank the anonymous reviewers for their thoughtful and constructive feedback on an early version of this work. 
One reviewer’s detailed critique helped crystallize the framing and motivated several of the conceptual refinements developed in the final discussion and epilogue.
G.M. is also grateful to D.A. for supervision and guidance.
This work is supported by the National Research Foundation, Singapore, and A*STAR under its CQT Bridging Grant and by the EU HORIZON—Project 101080085 – QCFD.

\newpage
\bibliographystyle{unsrt}
\bibliography{references}
\newpage

\input{sections/99-appendix}

\end{document}

%% file: preamble.tex
\usepackage{amsmath, amssymb, amsthm, bm, mathtools}
\usepackage{graphicx, subcaption}
\usepackage[hidelinks]{hyperref}
\usepackage{booktabs}
\usepackage[font=small]{caption}
\usepackage{tikz}
\usepackage{xcolor}
\usepackage{amsmath}
\usepackage{amssymb}
\usepackage{makecell}
\usepackage{amsthm}
\theoremstyle{definition}
\newtheorem{lemma}{Lemma}[section]
\newtheorem{proposition}{Proposition}[section]

\usepackage{needspace}
\usepackage{placeins}

\usepackage{pgfplots}
\pgfplotsset{compat=1.18}

\usetikzlibrary{positioning}
\usetikzlibrary{arrows.meta}
\usetikzlibrary{calc}

%% file: sections/00-abstract.tex
\begin{abstract}
    Qubit-efficient optimization studies how large combinatorial problems can be addressed with quantum circuits whose width is far smaller than the number of logical variables. In quadratic unconstrained binary optimization (QUBO), objective values depend only on one- and two-body statistics, yet standard variational algorithms explore exponentially large Hilbert spaces. We recast qubit-efficient optimization as a geometric question: what is the minimal representation the objective itself requires?
    Focusing on QUBO problems, we show that enforcing mutual consistency among pairwise statistics defines a convex body—the level-2 Sherali–Adams polytope—that captures the information on which quadratic objectives depend. We operationalize this geometry in a minimal variational pipeline that separates representation, consistency, and decoding: a logarithmic-width circuit produces pairwise moments; a differentiable information projection enforces local feasibility; and a maximum-entropy ensemble provides a principled global decoder.
    This information-minimal construction achieves near-optimal approximation ratios on large unweighted Max-Cut instances (up to $N=2000$) at shallow depth, indicating that pairwise polyhedral geometry already captures the relevant structure in this regime. By making the information-minimal geometry explicit, this work establishes a clean baseline for qubit-efficient optimization and sharpens the question of where genuinely quantum structure becomes necessary.
\end{abstract}

%% file: sections/01-intro-restructure.tex
\section{Introduction}\label{sec:intro}
\paragraph{The qubit bottleneck.}
Variational quantum algorithms (VQE \cite{peruzzo_variational_2014}, QAOA \cite{farhi_quantum_2014}) remain leading candidates for near-term combinatorial optimization.
Most advances pursue \textit{expressivity and trainability}—better ansatz \cite{zhu_adaptive_2022}, problem-informed mixers \cite{fuchs_constrained_2022}, warm-starting \cite{egger_warm-starting_2021}—motivated by the common intuition, summarized in \cite{tilly_variational_2022, cerezo_variational_2021}, that efficiently exploring a larger portion of Hilbert space may confer greater variational power.
Yet present-day quantum processors expose a simpler constraint: coherent circuit volume \cite{eisert_mind_2025}, the product of usable width and depth below the noise floor.
Although current hardware now hosts hundreds of physical qubits, only tens can participate in circuits of meaningful depth before decoherence dominates.
The practical limitation is therefore \textit{width}: how many variables can be represented coherently at once without inflating gate count or sampling overhead. This motivates a sharper question—\textbf{can we solve an $N$-variable optimization problem using fewer than $N$ qubits, without a compensating increase in circuit depth or total gate volume?}
This work addresses that question directly.

\paragraph{From expressivity to efficiency.}
When the number of qubits available is smaller than the number of optimization variables, three pragmatic routes emerge.
 One is to \textit{decompose} large instances into smaller subproblems that fit within hardware limits before recombining their partial solutions~\cite{maciejewski_multilevel_2024, acharya_decomposition_2024, bach_solving_2025}.
 A second is to \textit{compress}---encode multiple logical variables per qubit through shared bases or correlators, whether linearly \cite{fuller_approximate_2021}, polynomially \cite{sciorilli_towards_2025}, or exponentially  \cite{tan_qubit-efficient_2021}. A third is to leverage classical compute by embedding small quantum modules within \textit{hybridized} pipelines, where quantum correlations or relaxations improve classical preconditioning, rounding, or local-searching \cite{bach_solving_2025, dupont_optimization_2025, dupont_benchmarking_2025, tan_landscape_2024}. Each direction addresses the same width constraint from a different angle, and a working architecture may ultimately integrate all three. Here we focus on the \textit{compression} primitive: 
 the minimal representation of information that a quadratic objective requires.

\paragraph{Qubit-efficient compression schemes.}
Among the three routes above, \textit{compression} has evolved into a field of its own. It asks how many qubits are truly necessary to represent a combinatorial objective of $N$ variables. 
Early approaches drew inspiration from quantum random access codes (QRACs) \cite{fuller_approximate_2021}, which encode several classical variables probabilistically into the expectation values of a single qubit---achieving linear compression and providing a modest approximation ratio guarantee \cite{teramoto_quantum-relaxation_2023}. 
Contemporary to this, a line of qubit-efficient encodings \cite{tan_qubit-efficient_2021, huber_exponential_2024, leonidas_qubit_2024} achieved exponential width reduction by encoding multiple optimization variables within shared measurement bases, effectively capturing single-variable statistics while approximating pairwise correlations heuristically and conditionally. Between these extremes, Pauli Correlation Encoding (PCE) \cite{sciorilli_towards_2025} achieves polynomial compression, distributing logical variables across commuting Pauli correlators up to $N=8000$ variables of a QUBO problem. A complementary compression of the physical process such as the qubit-efficient QAOA \cite{sundar_qubit-efficient_2024} reduce qubit count by block-encoding the QAOA Hamiltonian itself. 
Together, these studies reveal that width can be reduced by orders of magnitude--but also sharpen a deeper question: \textbf{if linear, polynomial, and exponential compression are all possible, how much is enough?} 

For concreteness and tractability, both prior work and the present study limit this analysis to quadratic unconstrained binary optimization (QUBO) problems, where costs and correlations are well defined and theoretical comparisons are possible; the same reasoning is expected to generalize to broader combinatorial models.

\paragraph{Our contributions.}

The question for qubit-efficient optimization is not what a quantum system \textit{can} encode, but what it \textit{should}: what information the objective itself demands.
We answer this by identifying the information-minimal geometry underlying variational optimization for quadratic objectives. Focusing on QUBO problems, we show that the only statistics required to evaluate the objective are one- and two-body moments, and that enforcing their mutual consistency defines a well-characterized convex body: the Sherali–Adams level-2 polytope. 

We operationalize this geometry as an end-to-end variational pipeline: a logarithmic-width circuit produces pseudo-moments; a single $\rho$-damped iterative-proportional-fitting (IPF) step softly projects them toward local feasibility; and a maximum-entropy Gibbs ensemble provides a principled global decoder. This separation of representation, consistency, and decoding yields a minimal, differentiable framework that is quantum-compatible yet requires no non-commuting observables or positive-semidefinite (spectrahedral) constraints. Empirically, this information-minimal construction achieves near-optimal approximation ratios on large unweighted Max-Cut instances at low depth. By resisting the temptation to dress a minimal construction in additional structure, we isolate a parsimonious explanation for qubit-efficient optimization and, in doing so, sharpen where genuinely quantum geometry must enter.

\paragraph{Empirical performance.}

 On five \textsc{GSET} Max-Cut instances ($N=800$–$2000$), depth-2–3 circuits using a real-amplitude, hardware-efficient ansatz achieve best approximation ratios exceeding $0.99$; Erdős–Rényi graphs of varying density exhibit the same rapid saturation with depth. As detailed in Section~\ref{sec:experiments}, this behavior suggests that the minimal pairwise geometry already captures the relevant structure in this regime.

% \paragraph{Empirical low-depth performance.}
% Having established the geometric framework, we now examine its empirical performance.
% Across Erdős--Rényi graphs of varying densities and five \textsc{GSET} Max-Cut instances ($N{=}800$--$2000$),
% two layers of a real-amplitude hardware-efficient ansatz consistently achieve near-optimal performance:
% in four of five GSET cases the best approximation ratio exceeds $0.99$ at depth~2, 
% while the remaining instance (\textsc{G35}, $N{=}2000$) reaches $0.99$ at depth~3.
% These ratios are at or above the classical hardness thresholds 
% ($0.878$ for Goemans--Williamson; $16/17{\approx}0.941$ for the PCP inapproximability bound, 
% used as a reference benchmark in prior qubit-efficient studies~\cite{sciorilli_towards_2025}),
% indicating that the pairwise geometry saturates at shallow depth.
% Median results, reported in Appendix~\ref{app:rho-depth-facets}, follow the same trend.
% With this plateau established, any marginal gains from adding more quantum structure—such as increased depth, non-commutation, or complex amplitudes—are not only costly but must now be justified against simpler classical refinements of the geometric framework itself.
% By framing qubit-efficient optimization as geometry rather than heuristics, we can now ask not \emph{how} to make these algorithms more quantum, but \emph{when} quantum structure becomes necessary.

\section{From Global to Local: The Geometry of Quadratic Objectives}\label{sec:geometric_intuition}

\begin{figure}[h!]
    \centering
    \includegraphics[width=1.0\linewidth]{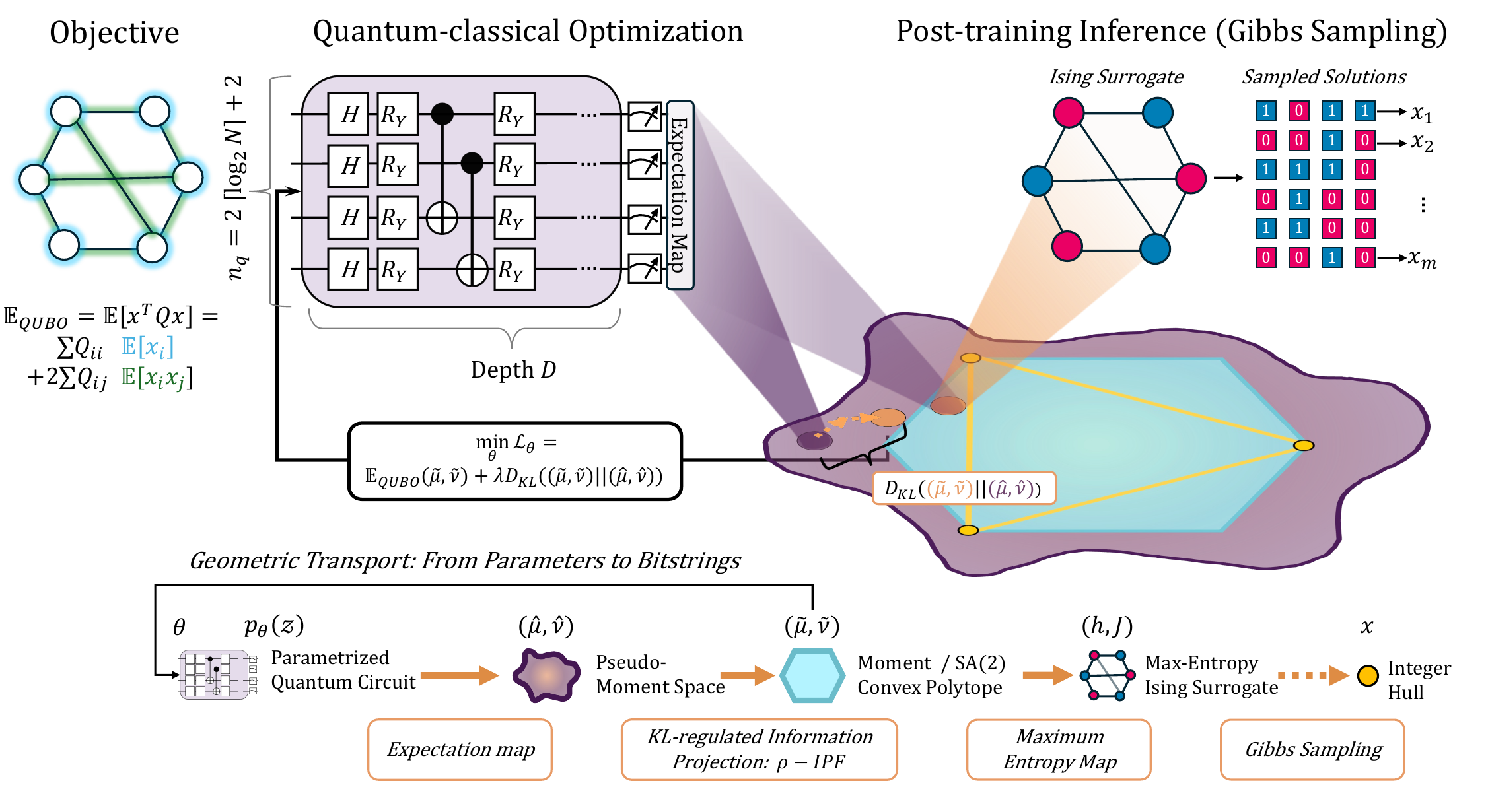}
    
    \caption{\textbf{The information-minimal two-body framework as geometric transport.}
    The framework treats optimization as a form of geometric transport: a sequence of maps that change representation while preserving the information relevant to the objective.
    \textbf{(Left)} A quadratic objective is determined by one- and two-body moments; a log-width circuit ($n_q = 2\lceil\log_2 N\rceil{+}2$) produces pseudo-moments $(\hat\mu,\hat\nu)$.
    \textbf{(Center)} A $\rho$-damped IPF step projects these toward the SA(2) polytope, yielding near-feasible $(\tilde\mu,\tilde\nu)$.
    \textbf{(Right)} A maximum-entropy Ising surrogate with fields $(h,J)$ enables Gibbs sampling of bitstrings.}
        \label{fig:main-pipeline}
\end{figure}

Intuitively, a quadratic objective on $N$ binary variables contains $\mathcal{O}(N^2)$ pairwise terms, since $Q\in\mathbb{R}^{N\times N}$ defines coefficients $Q_{ij}=Q_{ji}$.
Computing its expected value, $\mathbb{E}[x^\top Qx]$, depends only on expectations of single variables and their pairwise products—no higher-order correlations enter—although finding parameters that minimize this expectation remains an NP-hard search over $2^N$ configurations. The Hilbert space of an $N$-qubit circuit, by contrast, spans all $2^N$ amplitudes—far more information than is needed to represent or evaluate a quadratic objective, which depends only on these low-order correlations.
Leveraging this observation, we can focus training on a compact space of consistent local statistics rather than on the full state vector.
We now develop this idea by expressing the objective directly in this reduced set of expectations and treating their consistency as a geometric constraint.

\paragraph{The global view.} In a variational algorithm, the cost is normally written as a \emph{global} expectation 
\begin{align}
    \mathbb{E}_p[f]= \sum_{x} p(x) x^T Q x
\end{align}
or, in quantum form, $\braket{\psi| H_Q|\psi} = \sum_x | \braket{x|\psi}|^2 x^T Q x$,  where $p(x) = |\braket{x | \psi}|^2$ is the Born probability of bitstring $x$. The optimization seeks the circuit parameters that minimize this expectation value, then samples bitstrings from the resulting state—ideally obtaining the lowest-energy configuration with the highest probability.

While this \textit{global view} is natural for sampling, the same quantity can be expressed \textit{locally}.
\begin{align}
    \mathbb E_{p}[f]
=\sum_i Q_{ii}\,\mathbb E[x_i]
+2\sum_{i<j}Q_{ij}\,\mathbb E[x_i x_j]
=\sum_i Q_{ii}\mu_i+2\sum_{i<j}Q_{ij}\nu_{ij},
\end{align}
The two expressions are equivalent in value but distinct in perspective—the first averages over the full joint distribution $p(x)$, whereas the second depends only on the \textbf{local marginals} or moments, which we write $(\mu,\nu)=(\mathbb E[x_i],\mathbb E[x_i x_j])$. This local formulation is the one we build on: it exposes the minimal statistics that determine the objective and allows the consistency of those statistics to be treated geometrically.

\paragraph{The local view.} Moving from the global distribution $p(x)$ to its local statistics seems effortless---given all $2^N$ probabilities, the corresponding moments $(\mu,\nu)$ are trivial to compute. 
The reverse direction, however, is far from free: not every collection of local descriptors defines a valid global distribution. 
Compressing from $2^N$ probabilities to $O(N^2)$ local moments introduces feasibility constraints that ensure these numbers can coexist within one consistent probabilistic model. 
For binary variables the admissible region of $(\mu,\nu)$ follows directly from
non-negativity of the local $2\times2$ joint table for each pair $(i,j)$:
\begin{align*}
    P^{(ij)} =
    \begin{pmatrix}
    1-\mu_i-\mu_j+\nu_{ij} & \mu_j-\nu_{ij}\\[2pt]
    \mu_i-\nu_{ij}         & \nu_{ij}
    \end{pmatrix}.
\end{align*}
where $P^{(ij)}_{ab}=\Pr[X_i=a,\,X_j=b]$ for $a,b\in\{0,1\}$ (rows correspond to $x_i$, columns to $x_j$ in lexicographic order $00,01,10,11$). This is also known as the Venn diagram of probability, see Fig \ref{fig:venn-bf} of Sec. \ref{subsec:psuedo-to-feasible}.
Enforcing non-negativity yields the inclusion–exclusion (Boole–Fréchet) bounds.
\begin{align}\label{eq:boole-frechet}
    \max\{0,\,\mu_i+\mu_j-1\}\le \nu_{ij}\le \min\{\mu_i,\mu_j\}.
\end{align}

These inequalities simply express that joint probabilities cannot be negative.
For example, if $\mu_i=0.8$ and $\mu_j=0.6$, the joint moment $\nu_{ij}=P^{(ij)}_{11}$ must lie in $[0.4,0.6]$; values outside this bound would violate feasibility. 

\paragraph{Operationalizing Sherali--Adams.} This same pairwise geometry is exactly what the Sherali–Adams hierarchy formalizes at its second level~\cite{sherali_hierarchy_1994}.
Originally introduced as a lift-and-project hierarchy for analyzing the strength of polynomial-time relaxations, it tightens them by introducing higher-order variables and linear consistency constraints among them.
Although $\mathrm{SA}(2)$ has known integrality gaps and does not, in theory, close the NP barrier, it represents the tightest linear relaxation expressible using only one- and two-body correlations \cite{charikar_integrality_2009}.
Recent work~\cite{odonnell_sherali--adams_2018, odonnell_sherali--adams_2021} shows that low levels of the hierarchy can be surprisingly effective in practice—often yielding stronger empirical performance than asymptotic guarantees suggest.

Here we make $\mathrm{SA}(2)$ explicit and operational: the same mathematical object that underlies those theoretical results is treated as a concrete geometric constraint anchoring our optimization framework.
The convex region defined by these constraints coincides with the level-2 Sherali–Adams relaxation—the same geometry that guarantees a valid probabilistic model also provides the tightest two-body linear bound on a QUBO.
Anchoring the circuit to this convex body provides a principled baseline: every quantity learned by the model has a precise position within the known geometry of pairwise feasibility.
We give the full derivation and the equivalence to $\mathrm{SA}(2)$ in Sec.~\ref{subsec:sa2-bf-equivalence} and App.~\ref{app:sa2}.
This coincidence justifies using $\mathrm{SA}(2)$ as the canonical two-body target for learning and repair, and forms the geometric foundation for the projection and decoding steps developed in the next section.

\paragraph{Resolving the local-consistency gap.} In practice, the circuit’s outputs---its estimated moments $(\hat\mu,\hat\nu)$---will not in general satisfy these feasibility constraints exactly. 
This defines the \emph{local-consistency problem}: optimization can drift toward combinations of moments that are algebraically favourable but probabilistically unrealizable. 
To address this, we introduce two lightweight operations that complete the framework built on the $\mathrm{SA}(2)$ geometry. 
A single $\rho$-damped iterative-proportional-fitting (IPF) update ($\rho<1$) softly projects the pseudo-moments toward the feasible surface, maintaining differentiability for learning (Bregman form; App.~\ref{app:ipf} and~\cite{heinz_h_bauschke_method_1997}). 
The resulting locally-consistent marginals define a maximum-entropy Gibbs ensemble—the canonical way to sample bitstrings from the global distribution implied by the learned local statistics ~\cite{geman_stochastic_1984, koller_probabilistic_2010}.
In this view, Gibbs sampling serves as a \textit{distributional decoder}, producing a faithful distribution over integral solutions rather than relying on heuristic rounding or ad-hoc sampling schemes.
Their implementation details and empirical behavior are elaborated in the following sections.

These two operations—IPF for enforcing local consistency and Gibbs sampling for decoding a global distribution—directly address the challenges that earlier qubit-efficient models encountered. 
In Tan \emph{et al.}~\cite{tan_qubit-efficient_2021}, the absence of an explicit consistency rule meant that feasibility could only be guaranteed on $k$-regular or disjoint graphs, where local correlations did not overlap and could be averaged safely; larger, irregular instances had to be decomposed into small subgraphs to remain valid.
Pauli Correlation Encoding (PCE)~\cite{sciorilli_towards_2025} scaled to large irregular graphs, but enforced consistency implicitly through a tailored regularizer and a $\tanh$ non-linearity that approximates a probabilistic sign decoder for bistrings. 
Within our framework, these constraints are resolved explicitly and minimally: IPF supplies the minimal differentiable consistency correction, and Gibbs sampling provides a principled distributional decoder, completing in explicit form the mechanisms that prior schemes had to engineer implicitly.

\paragraph{Roadmap: The geometric transport.}
The decomposition of $x^\top Qx$ into one- and two-body statistics defines the sequence of geometric maps illustrated in Fig.~\ref{fig:main-pipeline}: 
amplitudes $\!\to\!$ pseudo-moments $\!\to\!$ repaired moments $\!\to\!$ an ensemble of bit-strings. 
The sections that follow examine these transports in order—reading moments directly from a log-width circuit, 
projecting and regularizing them toward pairwise feasibility within $\mathrm{SA}(2)$ by a KL-regulated information projection, 
and realizing the resulting convex geometry as a maximum-entropy Ising ensemble. Section \ref{sec:approach-framework}  outlines the geometry at a high level; Section \ref{sec:methods} then formalizes each transport and its numerical realization.

%% file: sections/02-approaches-rev.tex
\section{Our Approach: the Informational-Minimal Two-Body Framework}
\label{sec:approach-framework}
The framework follows directly from the sequence of geometric transports outlined Fig.~\ref{fig:main-pipeline} and Section~\ref{sec:intro}.
Each stage acts as a projection onto the next geometric surface:
expectations map amplitudes to moments,
a lightweight KL-based correction moves those moments toward convex feasibility,
and the resulting feasible geometry defines both the objective function value
and the Gibbs ensemble from which bit-strings are drawn.

\subsection{From Parameters to Pseudo-moments}
\label{subsec:parameters-to-pseudomoments}

The introduction established that every quadratic objective depends only on the
first- and second-order moments $(\mu,\nu)$ of the Born distribution $p_\theta(x)$.
We now describe the concrete statistics our circuit will produce.

We index variables by two address registers $(I,J)$ and read two value bits $(A,B)$.
Conditioning on the addressed pair yields the local probabilities
\[
\hat\mu_i := \Pr_\theta[X_i{=}1 \mid I{=}i \ \text{or}\ J{=}i],\qquad
\hat\nu_{ij} := \Pr_\theta[X_i{=}1,X_j{=}1 \mid \{I,J\}=\{i,j\}] .
\]
Because each pair ${i,j}$ is normalized independently, the mapping $p_\theta \mapsto \hat m = (\hat\mu,\hat\nu)$ is nonlinear yet fully differentiable, providing a smooth channel from amplitudes to local statistics.
The resulting quantities are \emph{pairwise} consistent—each edge defines a valid
$2{\times}2$ table—yet they need not be \emph{jointly} feasible across all edges.
We therefore refer to $(\hat\mu,\hat\nu)$ as \emph{pseudo-moments}:
locally valid marginals that may not correspond to a single global distribution.

Optimizing the linear energy $E_Q(\hat\mu,\hat\nu)$ directly can exploit such
inconsistencies by moving toward algebraically favorable but unrealizable
combinations (e.g.\ $\mu_i,\mu_j\!\to\!1$ with $\nu_{ij}\!\to\!0$ on an
anti-ferromagnetic edge). To prevent this drift while preserving gradients,
we introduce in §\ref{subsec:psuedo-to-feasible} a deterministic,
under-relaxed correction that gently guides the statistics \emph{towards}
pairwise feasibility during training.

\medskip
\noindent\emph{Implementation note.} The precise circuit realization is secondary; we employ a hardware-efficient ansatz that provides a smooth map $\theta \mapsto p_\theta(x)$ and, via the Born rule, to the local statistics above. Section~\ref{sec:moment-encoding} details the construction; here we focus on the geometry of these statistics and their regularization.

\subsection{From Pseudo-Moments Towards Locally Feasible Moments}
\label{subsec:psuedo-to-feasible}
We call these statistics \emph{pseudo-moments} because, if left unchecked,
the optimizer can exploit combinations that minimize the algebraic objective
while violating basic probability constraints.
Even when each edgewise table is internally valid,
the collection of tables can be \emph{locally inconsistent}—%
their shared marginals disagree on overlapping nodes—%
so that no single joint distribution can realize them simultaneously.
This distinction separates two notions of consistency.
\emph{Local consistency} ensures that each pairwise table obeys the Boole–Fréchet bounds;
\emph{global consistency} requires that all such tables agree on their overlaps.
A simple contradiction illustrates the gap:
if one pair $(i,j)$ yields $\nu_{ij}\!=\!1$ while another $(i,k)$ gives $\nu_{ik}\!=\!0$,
then the implied marginals for $X_i$ conflict—
in the first case $\mu_i$ must be~1, in the second~0.
A naïve reconciliation would assign $\mu_i\!=\!0.5$ everywhere, restoring feasibility but destroying information.
The challenge, therefore, is to recover consistency while retaining as much
structure as possible.

At the extreme, a configuration with $\mu_i\!\to\!1$, $\mu_j\!\to\!1$,
and $\nu_{ij}\!\to\!0$ is attractive whenever $Q_{ii},Q_{jj}\!<\!0$
and $Q_{ij}\!>\!0$, for it lowers the objective but corresponds to no valid
joint distribution.
Such points lie far outside the pairwise feasible region—the convex body of
admissible moments—often introduced in probability as the
\emph{Boole--Fréchet box} (Sec.~\ref{sec:intro}).
Figure~\ref{fig:venn-bf} shows its familiar Venn diagram form: a
$2{\times}2$ table of joint probabilities for two Bernoulli events
$A$ and $B$, where non-negativity of all four regions enforces the
inclusion–exclusion bounds that define the feasible interval for~$\nu_{ij}$.

\begin{figure}[h]
\centering
\begin{tikzpicture}[scale=1.0, every node/.style={font=\normalsize}]
% Background
% Background - outline only, no fill
\draw[gray!30] (-3.2,-1.8) rectangle (3.2,2.3);

% Circles
\begin{scope}
  \clip (-0.7,0) circle (1.3);
  \fill[blue!15] (0.7,0) circle (1.3);
\end{scope}
\draw[line width=1.2pt,blue!70] (-0.7,0) circle (1.3);
\draw[line width=1.2pt,red!70] (0.7,0) circle (1.3);

% Set labels at top
\node[blue!70,font=\Large] at (-0.7,1.6) {$A$};
\node[red!70,font=\Large] at (0.7,1.6) {$B$};

% Marginal probability labels
\node[blue!70,font=\normalsize] at (-1.3,0) {$\mu_i - \nu_{ij}$};
\node[red!70,font=\normalsize] at (1.3,0) {$\mu_j - \nu_{ij}$};

% Center: joint moment with equivalences
\node[font=\normalsize,align=center] at (0,0.0) {$\nu_{ij}$};

\node[font=\normalsize,align=center] at (0,-1.6) {$1 - \mu_i - \mu_j + \nu_{ij}$};

% Legend box
\node[anchor=center,align=center,font=\small,fill=white,fill opacity=0.01,
      text opacity=1,inner sep=4pt,rounded corners] at (0,-
    2.3) {
  $\mu_i = \Pr(A) = \mathbb{E}[X_i]$ 
  $\mu_j = \Pr(B) = \mathbb{E}[X_j]$ 
  $\nu_{ij} = \Pr(A{\cap}B) = \mathbb{E}[X_iX_j]$
};
\end{tikzpicture}

\caption{\textbf{Pairwise moment geometry.}
By a "$2{\times}2$ table" we mean the classical Venn-diagram representation of two Bernoulli events. 
For binary variables $X_i,X_j\!\in\!\{0,1\}$, 
$\mu_i=\Pr(A)=\mathbb{E}[X_i]$, $\mu_j=\Pr(B)=\mathbb{E}[X_j]$, and 
$\nu_{ij}=\Pr(A\cap B)=\mathbb{E}[X_iX_j]$ correspond to the regions shown. 
Nonnegativity of all four areas—$p^{ij}_{10} =(A\setminus B)$, $p^{ij}_{01} =(B\setminus A)$, $p^{ij}_{11} = (A\cap B)$, and its complement, $p^{ij}_{00}$—yields the inclusion–exclusion (Boole–Fréchet) bounds 
$\max\{0,\mu_i+\mu_j-1\}\le\nu_{ij}\le\min\{\mu_i,\mu_j\}$.}
\label{fig:venn-bf}
\end{figure}

\paragraph{A local repair for global feasibility.}
\textit{To restore and regularize feasibility while retaining structure,} we introduce a repair mechanism that gradually guides the statistics back toward the feasible region while keeping optimization smooth. 
We motivate it not by its form but by the properties it must satisfy: it should be \emph{fast}, relying only on local updates rather than a global solve; \emph{convex} in its corrective step, ensuring stability and a unique direction of improvement; and \emph{differentiable}, so gradients remain informative even when the moment is far from feasibility. 
A method that meets these criteria is the under-relaxed Iterative Proportional Fitting (IPF) update~\cite{heinz_h_bauschke_method_1997}.

As a simple illustration, take $(\mu_i,\mu_j,\nu_{ij})=(0.1,0.1,0.8)$ 
with a damping factor $\rho=0.6$.
The joint moment lies far outside its feasible interval 
$[L_{ij},U_{ij}]=[0,0.1]$, so the $\nu$-snap gives
\begin{align}
\nu_{ij}^{+}
   &= (1-\rho)\,\nu_{ij} + \rho\,U_{ij}
      &&= 0.4\times0.8 + 0.6\times0.1 = 0.38. \nonumber
\end{align}
The updated joint raises the lower bound for each marginal to 
$L_i=\max_j\nu_{ij}^{+}=0.38$
and sets the upper bound 
$U_i=\min_j(1+\nu_{ij}^{+}-\mu_j)=1+0.38-0.1\approx1$.
Projecting $\mu_i$ into this interval and damping gives
\begin{align}
\mu_i^{+}
   &= (1-\rho)\,\mu_i + \rho\,\Pi_{[L_i,U_i]}(\mu_i)
      &&= 0.4\times0.1 + 0.6\times0.38 = 0.268. \nonumber
\end{align}
By symmetry $\mu_j^{+}=0.27$, so after one pass
\[
(\mu_i^{+},\mu_j^{+},\nu_{ij}^{+})\approx(0.268,0.268,0.38),
\]
a smooth contraction toward the Boole--Fréchet region that 
reduces infeasibility while preserving gradient signal. 
In higher dimensions, these local ``snaps'' interact across edges, 
producing coupled dynamics, but this toy example captures the essential behavior.

Among heavier alternatives, belief propagation \cite{koller_probabilistic_2010} restores global consistency
through message passing on factor graphs but is inexact on loopy structures,
while optimal-transport \cite{cuturi_sinkhorn_2013} schemes compute the shortest feasible displacement under an information metric at substantial computational cost.
The IPF update occupies the middle ground: it is convex, differentiable,
and monotone, guiding pseudo-moments toward feasibility rather than solving
a nested optimization for the optimal path.
This light machinery is sufficient for our purpose—the geometry need only
be approached consistently, not reached optimally.

The under-relaxation ensures monotonicity: each update reduces the total
violation of the constraints without snapping the statistics abruptly onto
the boundary. This yields a deterministic map
$\hat{m}\!\mapsto\!\tilde{m}$ that moves each edgewise table toward the nearest feasible point under KL divergence—the natural distance measure
between probability or moment distributions.
Far from the feasible region, the update acts as a projection and penalizes large corrections; near the boundary, it transitions into a soft regulator that prevents drift and maintains smooth gradients.

The accompanying relative-entropy term
$\mathrm{KL}(\tilde{m}\|\hat{m})$ makes the correction cost explicit:
pseudo-moments that lie deeper outside $\mathrm{SA}(2)$ incur a larger
information-theoretic penalty.
Coupled to the training objective, it encourages the circuit to generate
naturally feasible outputs rather than rely on post-hoc repair. Formally, this step is a Bregman projection under the negative-entropy potential,
whose analytic form we detail in Sec.~\ref{app:ipf}.
Together, the damped update and the KL penalty stabilize training so that
feasibility is approached gradually, and the geometry of the objective
remains coherent throughout the optimization.

The following section formalizes this intuition as a loss function
$\mathcal{L}(\theta)$ combining the projected QUBO energy with a KL regularization
term which penalizes deviations from feasibility.

\textit{In practice,} an intermediate damping value performs best: $\rho\!\approx\!0.5$ consistently stabilizes training, while both extremes—no correction ($\rho{=}0$) and hard projection ($\rho{=}1$)—tend to degrade performance, producing a characteristic U-shaped behavior in the loss (see Sec.~\ref{subsec:ER-results-dyanmics}).

\subsection{One Polytope, Two Languages}\label{subsec:sa2-bf-equivalence}

What probability calls the Boole--Fréchet bounds, optimization knows as the Sherali--Adams hierarchy. At level 2 of the hierarchy, they describe exactly the same convex body. The set of all $(\mu, \nu)$ that satisfy the inclusion--exclusion inequalities forms the pairwise feasible region
\begin{align*}
\mathcal{P}_{\text{pair}} = \{(\mu, \nu) : 0 \le \mu_i \le 1, \; \max\{0, \mu_i + \mu_j - 1\} \le \nu_{ij} \le \min\{\mu_i, \mu_j\} \; \forall i \neq j\}.
\end{align*}
In the Sherali--Adams construction, a 0--1 integer program is relaxed by introducing new variables for pairwise products $y_{ij} = x_i x_j$ and imposing the linear bounds $0 \le y_{ij} \le x_i$, $0 \le y_{ij} \le x_j$, $y_{ij} \ge x_i + x_j - 1$. Identifying $x_i \leftrightarrow \mu_i$ and $y_{ij} \leftrightarrow \nu_{ij}$ shows that these are exactly the Boole--Fréchet inequalities that define $\mathcal{P}_{\text{pair}}$; the short derivation appears in Appendix \ref{app:sa2}. In compact form,
\begin{align}
\mathcal{P}_{\text{pair}} \equiv \mathrm{SA}(2) \qquad \text{for QUBO.}
\end{align}
Level 2 of the Sherali-Adams hierarchy is the tightest convex relaxation obtainable from first- and second-order moments alone \cite{sherali_hierarchy_1994}. This coincidence explains both our representation and our notation: the quantities $(\mu, \nu)$ are moments in the \emph{statistical} sense---expectations of the Born distribution---and simultaneously the moment variables of \emph{optimization theory}. 

In this shared geometry, probability and optimization meet on the same convex surface. The consistency term we enforce is, therefore, not an invented penalty but the requirement that the circuit's learned statistics remain within a convex body already known to capture the optimal two-body relaxation of the discrete problem. One polytope, two languages.

\subsection{From Moments to the Ensemble}

The learned moments encode a distribution; we now sample from it. Among all distributions consistent with these marginals, the one of maximum entropy assumes nothing beyond them. It takes the familiar Ising form
\begin{align*}
p(x) \propto \exp\!\left(\sum_i h_i x_i + \sum_{i<j} J_{ij} x_i x_j\right),
\end{align*}
with parameters $(h, J)$ fixed by log-odds relations (see Sec. \ref{sec:ising-surrogate}).

Sampling from this model through Gibbs updates converts the continuous
geometry of moments into a discrete ensemble of bit-strings, extending local
pairwise consistency to the full graph.
The resulting ensemble is probabilistically consistent with the learned
marginals: together, its members realize, in discrete form, the statistics
encoded by the point in moment space.

Among possible decoders, the Gibbs sampler is the natural choice—it is the
simplest method for drawing from the global maximum-entropy distribution
implied by those local marginals.
No further optimization is performed; the sampler merely expresses the
geometry the circuit has already learned.
Both the IPF repair and the Gibbs decoder are classical in machinery but
geometric in function: one projects, the other samples.
Neither evaluates nor minimizes the true objective; together they complete the
transport from moments to an ensemble, where the learned geometry becomes
a distribution over discrete solutions.

%% file: sections/03-methods.tex
% sections/03-methods.tex
\section{Methods}\label{sec:methods}

The preceding section described the framework conceptually as a sequence of geometric maps. Here we specify its concrete realization: a log-width circuit producing pseudo-moments, a lightweight IPF-based correction that regularizes them toward feasibility, and a Gibbs decoder sampling from the resulting maximum-entropy ensemble.

\input{sections/methods/03a-moment-encoding-staging}
\input{sections/methods/03b-sa2_ipf_projection}
\input{sections/methods/03c-ising_surrogate}
\input{sections/methods/03d-training_and_implementation}

%% file: sections/methods/03a-moment-encoding-staging.tex
\subsection{Pairwise Moment Readout with Log-width Address Registers}
\label{sec:moment-encoding}

We now describe how the circuit encodes pairwise moments with logarithmic width.
The Born distribution $p_\theta(i,a,j,b)=|\alpha_{iajb}(\theta)|^2$
defines the joint statistics over addresses $(I,J)$ and binary readouts $(A,B)$,
from which we obtain the one- and two-body moments $(\mu,\nu)$ used
throughout the framework.

We find this geometric and probabilistic formulation the cleanest way to reason about the framework. Readers who prefer the same construction expressed in projectors, operators, and traces will find the equivalent quantum version in Appendix~\ref{app:quantum-version-of-moments}.

\paragraph{Register structure.}
We use four registers ordered as $(I,A,J,B)$ with
$(I,J)\in[N]\times[N]$ and $(A,B)\in\{0,1\}^2$.
The circuit prepares
\[
\ket{\Psi(\theta)} 
= \sum_{i,j=1}^N \sum_{a,b\in\{0,1\}} \alpha_{i a j b}(\theta)\, \ket{i,a,j,b},
\]
with Born probabilities $p_\theta(i,a,j,b)=|\alpha_{i a j b}(\theta)|^2$.
We write $p_{i a j b}$ for short.

Thus, an $N\times N$ instance requires $n_q=2(\lceil\log_2 N\rceil+1)$ qubits under our scheme.

\paragraph{Hardware-efficient ansatz and layer count.}
Following the hardware-efficient ansatz (HEA) of~\cite{kandala_hardware-efficient_2017}, 
we use a layered circuit composed of single-qubit $R_y$ rotations and fixed entangling stages of CNOT gates.
Each \emph{HEA layer} consists of one parallel $R_y$ rotation on all qubits followed by two brickwork layers of CNOTs acting on disjoint qubit pairs
$(0,1),(2,3),\ldots$ and $(1,2),(3,4),\ldots$, respectively.
An initial Hadamard layer prepares the uniform superposition $\ket{+}^{\otimes n_q}$.
We denote by $D$ the number of such HEA layers,
so that the total physical gate depth is approximately $3D{+}1$,
accounting for the initial Hadamards and the three sublayers ($R_y$, even-CNOT, odd-CNOT) per block.

\paragraph{Observation events.}

We regard $I,J$ as address registers selecting which variables to observe, and $A,B$ as their binary readouts.
Define the observation events
\begin{align*}
    \mathcal{O}_i \coloneqq \{(I=i,J=j): j\neq i\}\ \cup\ \{(I=j,J=i): j\neq i\},\qquad
\mathcal{O}_{\{i,j\}} \coloneqq \{(I=i,J=j)\}\ \cup\ \{(I=j,J=i)\},
\end{align*}

excluding self-pairs $(I{=}J{=}i)$.
\emph{We do not preselect which $(i,j)$ to observe}: the addresses $(I,J)$ are drawn from the circuit’s Born
distribution, and the conditional definitions below remove any selection bias.

\paragraph{Moments as conditional expectations.}
Moments are computed as conditional probabilities:
\begin{align}
  \boxed{\
  \mu_i = \Pr_\theta[x_i{=}1 \mid i\text{ observed}],\qquad
  \nu_{ij} = \Pr_\theta[x_i{=}1,x_j{=}1 \mid \{i,j\}\text{ observed}]\
  }  
\end{align}

Here $x_i$ denotes the logical bit associated with decision variable $i$; the circuit encodes information for this variable in the ancilla value $A$ (or $B$) when the corresponding address $I= i ($ or $J = i)$ is sampled.

Since $x_i\in\{0,1\}$, probabilities equal expectations for Bernoulli variables, so equivalently
\begin{align*}
    \mu_i=\mathbb{E}_\theta[x_i\mid \mathcal{O}_i],\qquad
    \nu_{ij}=\mathbb{E}_\theta[x_i x_j\mid \mathcal{O}_{\{i,j\}}].
\end{align*}

That is, $\mu_i$ is the chance the addressed bit $i$ equals $1$, and $\nu_{ij}$ is the chance both addressed bits $i$ and $j$
equal $1$ when the unordered pair $\{i,j\}$ is observed in any order.

\paragraph{Moment formulas.}
Aggregating contributions from both address orders and excluding self-correlations, we obtain
\begin{equation}
\mu_i =
\frac{\displaystyle
   \sum_{j\neq i}\sum_{b} p_{i,1,j,b}
 + \sum_{j\neq i}\sum_{a} p_{j,a,i,1}}
{\displaystyle
   \sum_{j\neq i}\sum_{a,b}\!\big(p_{i,a,j,b}+p_{j,a,i,b}\big)},
\label{eq:mu}
\end{equation}
and, for $i\neq j$,
\begin{equation}
\nu_{ij} =
\frac{p_{i,1,j,1}+p_{j,1,i,1}}
     {\sum_{a,b}\!\big(p_{i,a,j,b}+p_{j,a,i,b}\big)},
\qquad
\nu_{ij}=\nu_{ji},\quad \nu_{ii}\coloneqq 0.
\label{eq:nu}
\end{equation}
These conditionals produce values in $[0,1]$ by construction.
Because they are computed under different conditioning events, we refer to $(\widehat\mu,\widehat\nu)$ as
\emph{pseudo-moments}—they need not be jointly feasible on the full graph (see §\ref{sec:sa2-ipf}).

%% file: sections/methods/03b-sa2_ipf_projection.tex
\subsection{Projection toward SA(2) via $\rho$-damped IPF}
\label{sec:sa2-ipf}

\paragraph{Pseudo-moments and feasibility.}
Raw circuit outputs $(\hat\mu,\hat\nu)$ are \emph{pseudo-moments}—provisional
estimates that need not satisfy pairwise consistency.
For statistical decoding to be valid, each edge $(i,j)$ must admit a nonnegative
$2{\times}2$ joint table, enforcing the Boole–Fréchet bounds:
\begin{align*}
    \max\{0,\mu_i+\mu_j-1\} \le \nu_{ij} \le \min\{\mu_i,\mu_j\}.
\end{align*}
The polytope defined by these constraints for all edges (together with symmetry
$\nu_{ij}{=}\nu_{ji}$ and $\nu_{ii}{=}0$) is precisely the Sherali–Adams level-2
(SA(2)) relaxation (§\ref{subsec:sa2-bf-equivalence}).
We project pseudo-moments toward SA(2) to ensure decodability while preserving
gradient flow. Even when each $2{\times}2$ table is locally valid, the family can disagree on overlaps; the IPF step reconciles these marginal inconsistencies while maintaining differentiability for variational learning.

\paragraph{Algorithm: $\rho$-damped IPF.}
We apply one iteration of damped iterative proportional fitting (IPF):

\noindent\textbf{(i) $\nu$-box snap.}
For each pair $(i,j)$, clip $\nu_{ij}$ to its Boole–Fréchet interval
\begin{align*}
    L_{ij}=\max\{0,\mu_i{+}\mu_j{-}1\}, \qquad
    U_{ij}=\min\{\mu_i,\mu_j\},
\end{align*}
and update
\begin{align*}
    \nu_{ij}^{\text{new}}
    =(1{-}\rho)\,\nu_{ij}
     +\rho\,\mathrm{clip}(\nu_{ij},L_{ij},U_{ij}),
\end{align*}
then symmetrize and set $\nu_{ii}{=}0$.

\noindent\textbf{(ii) $\mu$-consistency snap.}
Update each marginal $\mu_i$ using bounds implied by
$\nu^{\text{new}}$:
\begin{align*}
    L_i=\max_j \nu_{ij}^{\text{new}}, \qquad
    U_i=\min_j(1+\nu_{ij}^{\text{new}}-\mu_j),
\end{align*}
\begin{align*}
    \mu_i^{\text{new}}
    =(1{-}\rho)\,\mu_i
     +\rho\,\mathrm{clip}(\mu_i,L_i,U_i).
\end{align*}
The damping parameter $\rho\!\in\!(0,1]$ controls the projection strength:
$\rho{=}1$ applies a full correction, while smaller $\rho$ provides a softer
nudge toward feasibility.  We fix $\rho{=}0.5$ and perform a single iteration
($T{=}1$), sufficient to stabilize gradients without full convergence.
A more formal analysis of the algorithm's relation to gradients appears in
Appendix~\ref{app:ipf-followup}.
Furthermore, although we refer to this step as a “KL projection,” it is formally a Bregman projection under the negative-entropy potential (see App.~\ref{app:ipf}).

\paragraph{Training loss.}
We optimize the QUBO energy on the projected moments with a KL regularization term:
\begin{equation}
\boxed{\mathcal{L}(\theta)=E_{\text{QUBO}}(\tilde\mu,\tilde\nu)
+\lambda_{\text{KL}}(t)\,
\mathrm{KL}\!\big((\tilde\mu,\tilde\nu)\,\|\,
(\hat\mu,\hat\nu)\big)}
\end{equation}
where
$E_{\text{QUBO}}(\mu,\nu)
=\sum_i Q_{ii}\mu_i
 +2\sum_{i<j}Q_{ij}\nu_{ij}$.
The KL term penalizes large corrections, encouraging the circuit to produce
naturally near-feasible outputs.
The weight $\lambda_{\text{KL}}(t)$ is scheduled during training—
small early (prioritizing energy) and larger later (enforcing feasibility);
its analytic form and parameters are listed in
Appendix~\ref{app:ablation}.

Unless otherwise stated, we use $\rho{=}0.5$ and a single iteration ($T{=}1$); see Sec.~\ref{sec:training-impl} for the global training policy.

Limiting to a single under-relaxed pass also maintains the minimal classical
footprint of the method: additional inner loops would increase the effective
computational power of the classical side.
In principle, the effective projection strength depends jointly on $\rho$ and $T$
(roughly $1-(1{-}\rho)^T$),
so future work may explore smaller $\rho$ and larger $T$ as a smoothness–stability trade-off.

%% file: sections/methods/03c-ising_surrogate.tex
\subsection{Maximum-entropy Ising Surrogate and Gibbs Decoding}
\label{sec:ising-surrogate}

\paragraph{Decoding strategy.}
Having obtained the repaired moments $(\tilde\mu,\tilde\nu)$ , we next decode them into discrete bitstrings corresponding, ideally, to good solutions of the original objective.

$(\tilde\mu,\tilde\nu)$ characterizes a fractional point near or on $\mathrm{SA}(2)$—they specify pairwise local marginals but not a unique global distribution.
Rather than rounding (which discards the learned correlations), we adopt the maximum-entropy distribution consistent with these moments, corresponding to the exponential family whose sufficient statistics are the one- and two- body terms $x_i$, $x_i x_j$: the pairwise Ising model. Among all distributions with the given marginals, it has maximum entropy and \emph{admits a pairwise Markov random field (edge-factorized) representation}, making it the canonical choice …
 for probabilistic graphical models \cite{geman_stochastic_1984,koller_probabilistic_2010}.
Although computing the exact maximum-entropy distribution on loopy graphs is NP-hard, sampling from it is tractable: Gibbs updates require only local pairwise information—the same structure the model has already learned.

\paragraph{Moments to Ising Hamiltonian.}
Having obtained the repaired moments $(\tilde\mu,\tilde\nu)$—hereafter denoted
$(\mu,\nu)$ for simplicity—we decode them into discrete bitstrings. Let $\mu_i=\Pr[x_i{=}1]$ and $\nu_{ij}=\Pr[x_i{=}1,x_j{=}1]$ denote the (repaired) moments obtained after the $\rho$-damped projection. Form the $2{\times}2$ marginal table for each edge $(i,j)$:
\begin{align}
P^{ij}_{11}&=\nu_{ij}, &
P^{ij}_{10}&=\mu_i-\nu_{ij}, &
P^{ij}_{01}&=\mu_j-\nu_{ij}, &
P^{ij}_{00}&=1-\mu_i-\mu_j+\nu_{ij}.
\end{align}
Working in spins $\sigma\!\in\!\{-1,+1\}^N$ with $\sigma=2x{-}1$, the edge coupling
is given by the log-odds ratio:
\begin{align}
J_{ij} \;=\; \tfrac14 \log\!\left(\frac{P^{ij}_{11}P^{ij}_{00}}
                                     {P^{ij}_{10}P^{ij}_{01}}\right),
\qquad (i\neq j), \quad J_{ii}=0,
\end{align}
and the local fields are chosen to reproduce the magnetization
$m_i=2\mu_i{-}1$:
\begin{align}
h_i \;=\; \tfrac12 \log\!\frac{\mu_i}{1-\mu_i}
\;-\; \sum_{j} J_{ij} m_j.
\end{align}
This “edgewise” log-odds construction enforces each $2{\times}2$ marginal exactly and adjusts fields to match node means; on loopy graphs it is the standard closed-form MaxEnt embedding for pairwise MRFs. We do not claim that it solves the full global convex moment-matching problem; our goal here is a simple, principled decoder whose marginals coincide with the repaired two-body statistics on each edge.
To preserve sparsity and keep the per-sweep cost linear in edges, we
\emph{mask} $J$ to the nonzero pattern of $Q$—%
setting $J_{ij}{=}0$ wherever $Q_{ij}{=}0$—%
and recompute $h$ using the same expression.
We fix $\beta{=}1$, avoiding temperature tuning or heavier MCMC
machinery whose effects are largely well understood.
The goal is to keep the decoder minimal, attributing performance to the learned moments rather than to complex sampling dynamics.

In practice, the repaired moments may not satisfy all feasibility bounds exactly,
especially early in training.
To maintain numerical stability and ensure that all entries of $P^{ij}$ remain
positive, we clip each $P^{ij}_{ab}$ to a small floor
$\epsilon\!\approx\!10^{-12}$ before taking logarithms.
This regularization allows the sampler to operate reliably even when moment
estimates are imperfect, supporting early-training diagnostics without requiring
full feasibility.
We do not claim formal convergence of the combined IPF–Gibbs procedure to a
fixed point; our results show only that the learned statistics are stable in
practice and yield high-quality samples under consistent training settings. This mapping embeds the repaired moment point into an energy landscape whose
marginals coincide with those learned geometrically.

\paragraph{Gibbs sampling decoder.}
Single-site heat-bath updates are applied with local field $f_i=h_i+\sum_{j}J_{ij}\sigma_j$ and update probability
\begin{align}
\Pr[\sigma_i{\leftarrow}+1 \mid \sigma_{\setminus i}]
\;=\; \tfrac12 \big(1+\tanh(\beta f_i)\big).
\end{align}
Local fields are updated incrementally, so each sweep costs $O(|E|)$ work.
For sparse graphs this is linear in $N$.
In our runs we allocate $T=O(N\log N)$ sweeps per chain—a standard heuristic for the mixing time of sparse Ising systems—with exact sweep counts for the \textsc{GSET} problem class reported in Appendix~\ref{app:gset_tables}.
We run $n{=}8$ independent chains in parallel and return the best bitstring under the physical objective $E(x)=x^\top Qx$.
Because the chains evolve independently, the probability that none reach a satisfactory configuration decreases as $(1-\delta)^n$ where $\delta$ denotes the empirical success probability of a single chain within some target quality threshold. This rule-of-thumb argument simply motivates the use of parallel chains: the scheme is \emph{embarrassingly parallel}, lowering failure probability without increasing sequential runtime.

We use Gibbs purely as a \emph{sampling tool} parameterized by the learned
moments, not as an optimizer seeking a stationary distribution.
Unlike a classical optimizer, Gibbs has no access to or awareness of the
objective $E_{QUBO}(x)$; it samples configurations consistent with the learned
statistics rather than minimizing the cost directly.
This keeps attribution on moment quality rather than on stochastic search.

Our implementation uses \texttt{Numba} just-in-time (JIT) compilation to fuse update and reduction kernels; after warm-up, full \textsc{GSET} instances up to $N{=}2000$ complete tens of thousands of sweeps in only a few seconds.
Representative per-instance runtimes are provided in Appendix~\ref{app:gset_tables}, confirming that sampling cost is negligible relative to training.
All reported results use raw Gibbs samples with no additional local search or post-processing.

%% file: sections/methods/03d-training_and_implementation.tex
\subsection{Training and Implementation}\label{sec:training-impl}

Numerical experiments are implemented in \texttt{PyTorch} using a
hardware-efficient ansatz (all-qubit $R_y$ rotations followed by two
brickwork CNOT sublayers), yielding a two-qubit time depth of $2D$ at
logarithmic width $n_q = 2\lceil\log_2 N\rceil + 2$.
Born probabilities over $(I,A,J,B)$ are accumulated into
pseudo-moments $(\hat{\mu},\hat{\nu})$ via a single GPU
\texttt{scatter\_add}, followed by one $\rho$-damped KL/IPF projection
($T{=}1$) toward the $\mathrm{SA}(2)$ polytope; we then evaluate
$E_Q(\tilde{\mu},\tilde{\nu})$.
Training uses the \texttt{Adam} optimizer with learning-rate scheduling,
and decoding fits a maximum-entropy Ising model on the sparsity pattern of
$Q$ with parallel Gibbs chains. State-vector simulations were performed with single-precision values used throughout (\texttt{float32}) to balance numerical stability and computational efficiency.
Baselines cache of best-known values and state-of-the-art commercial solvers are computed (Bure-Monteiro, and Gurobi), and results are reported as \textbf{anytime} metrics—best-so-far
performance under equal budgets.
All runs use an \texttt{Apple M4 Pro} with Metal Performance Shaders (MPS)
for GPU acceleration.

\paragraph{Training regime.}
We operate in the analytic (statevector) regime:
Born probabilities $p_\theta(i,a,j,b)=|\alpha_{iajb}(\theta)|^2$ are
evaluated exactly, and gradients are obtained by automatic differentiation.
This regime enables efficient backpropagation and isolates the behavior of
the geometric representation from stochastic shot noise.
On hardware, the same quantities can be estimated empirically by averaging
finite shots over the observation events of
Sec.~\ref{sec:moment-encoding}; the learning formulation is identical.
Because the circuit width grows only logarithmically with problem size,
the resulting state space remains efficiently simulatable on classical
hardware.
This is intentional: the goal is not to demonstrate hardware dependence
but to test what performance is achievable when the geometry is made
explicit and the optimization is clean.
Across the representative instances studied in the next section, this
log-width, analytically trained model already achieves high empirical
performance.
Our approach follows a minimal-complexity principle: if such a circuit
already captures the essential structure of the problem, adding resources
merely to render it classically unsimulatable offers no explanatory gain.
The value of future quantum extensions therefore lies not in being harder
to simulate, but in revealing structure that the simplest geometry cannot.

\paragraph{Stabilization policies.}
A KL-penalty ramp prevents the feasibility term from dominating early in
training, allowing smooth convergence even when the initial moments are far
from $\mathrm{SA}(2)$.
The learning rate follows a linear warm-up–hold–exponential-decay schedule:
the warm-up phase gradually ramps Adam’s internal statistics and moves the
optimizer into a well-conditioned region before high-LR exploration, while
the exponential decay stabilizes convergence toward the end of training.
A peak learning rate of $\eta_{\mathrm{peak}}{=}0.10$ yields consistently
faster and more stable convergence than the default $0.01$ commonly used in
variational-circuit training.
Unless otherwise stated, all reported experiments use this fixed
combination of KL-penalty ramp and learning-rate schedule; only the
projection strength $\rho$, circuit depth $D$, and total epoch budget
(varying with $N$) are modified in the ablation and scaling studies.
Analytic forms and parameter values for both schedules are provided in
Appendix~\ref{app:ablation}.

%% file: sections/04-experiments.tex
\input{sections/experiments/04-experiments}

%% file: sections/experiments/04-experiments.tex
\section{Experiments and Results}
\label{sec:experiments}

\paragraph{Overview and baselines.}
We evaluate the $\mathrm{SA}(2)$-constrained two-body solver on unweighted
Max-Cut---a canonical NP-hard benchmark that is well characterized and
widely used to assess variational algorithms.
The goal is not to fine-tune per instance, but to test whether a single, geometry-constrained configuration performs consistently across random and
structured graphs and scales gracefully in~$N$.

We compare against three classical references:
\emph{Gurobi} (powerful commercial solver, 10\,min limit);
\emph{Burer--Monteiro \cite{burer_nonlinear_2003}} (low-rank semi-definite relaxation), evaluated in
two modes---BM-SS (single random start, as used in \cite{sciorilli_towards_2025}) and BM-MS (standard multi-start with
identical 10\, min limit) both use the implementation in \textsc{MQLIB} \cite{dunning_what_2018}; and
\emph{SA(2)-LP$\to$,Gibbs}, a linear-relaxation baseline that solves the $\mathrm{SA}(2)$ Linear Program (LP) formulation exactly—i.e., the standard Sherali–Adams level-2 relaxation of the QUBO—and then decodes the fractional solutions $(\mu_i,\nu_{ij})$ with the same Gibbs budget as our method.
These references jointly bracket the best-known practical range of
approximation ratios on \textsc{Gset} under comparable budgets.
For completeness, we also tested a direct QUBO$\!\to$Ising formulation decoded by Gibbs sampling.
Without temperature tuning the chain remains effectively ``cold,'' leading to poor mixing; we apply a robust scaling to obtain stable though weak results, reported in App.~\ref{app:gset_tables}. Because its performance is an order of magnitude worse, we omit it from the main figure for visual clarity.

All numerical experiments use the frozen policy identified in
Appendix~\ref{sec:ablation} through systematic hyper-parameter sweeps:
projection strength~$\rho{=}0.5$,
linear KL-penalty ramp~$0.10{\rightarrow}0.30$, and a
\texttt{warmup--hold--exponential} learning-rate schedule
($f_h{=}0.40$, $\eta_{\mathrm{peak}}{=}0.10$).
Training and sampling costs are reported in
Appendix~\ref{app:gset_tables} for $\textsc{GSET}$ instances.

\paragraph{Experimental design.}
Hyper-parameters were tuned exclusively on Erdős–Rényi (ER) random graphs~\cite{erdhos_random_1959}, 
chosen for their statistical neutrality and well-understood theoretical properties.
This ensemble offers a controlled setting in which average degree $\alpha=p(N{-}1)$ can be tuned directly, 
ensuring that observed trends reflect the solver’s geometry rather than artifacts of graph topology.
We sweep $\alpha\!\in\!\{1,4,5,8,12\}$, spanning three characteristic connectivity regimes:
(i) a \emph{tree-like} regime near the percolation threshold ($\alpha{\sim}1$),
(ii) a \emph{sparse, mesoscopic} or \emph{critical} regime ($\alpha{\approx}4$–$5$),
where small cycles and correlated motifs begin to appear, and
(iii) a \emph{moderately dense} regime ($\alpha{\in}\{8,12\}$),
where overlapping neighborhoods test higher-order consistency.

Each configuration is evaluated over multiple graph instances
(10 for $N{<}256$, 5 for $N{\ge}256$, for each $\alpha$), yielding several thousand training runs in total.
Across these sweeps we identified a single robust setting--
$\rho{=}0.5$ (IPF under-relaxation),
KL-ramp $0.10{\rightarrow}0.30$,
and a \texttt{warmup-hold-exponential decay} learning-rate schedule
($f_h{=}0.40$, $\eta_{\mathrm{peak}}{=}0.10$)---which we freeze for all
headline evaluations on \textsc{Gset} \cite{ye_yinyu_gset_nodate}.
Circuits are trained 
with \texttt{Adam}.

\paragraph{Reporting convention.}
For ER graphs we report the \emph{final gap} after a fixed training budget,
\[
\mathrm{gap}_{\mathrm{final}} = 1 - r_T, \qquad
r_T = \frac{\mathrm{cut}(\hat x_T)}{\mathrm{OPT}},
\]
to evaluate stability under a controlled budget.
For Gset we report the \emph{anytime best}, or \emph{incumbent} solution
\[
r^\star = \max_{t\le T}\frac{\mathrm{cut}(\hat x_t)}{\mathrm{OPT}},
\]
which reflects modern iterative-solver practice.
Sampling occurs every 30 epochs (and every 10 epochs in the final four intervals);
this cadence is intentionally conservative—denser decoding would only refine the
anytime curve but does not change the final ratio within the noise level of our runs.

For Erdős–Rényi experiments, we report the \emph{final gap} after a fixed budget because these runs serve primarily as controlled ablations. The absolute cut value is less important than the stability of training: the final gap reflects whether gradients converge smoothly, without spikes or collapse, and thus serves as a proxy for training reliability. In contrast, for structured \textsc{Gset} benchmarks we follow standard optimization practice and report the \emph{anytime best} (incumbent) ratio, since classical solvers also measure progress by the best feasible cut achieved within a computational budget.

For ER graphs, $\mathrm{OPT}$ denotes the best solution found by \emph{Gurobi}
within a 10-minute limit on each instance.
For \textsc{Gset}, $\mathrm{OPT}$ corresponds to the best-known cut value obtained
by the \emph{Breakout Local Search (BLS)} heuristic~\cite{benlic_breakout_2013},

\begin{figure}[!h]
\centering
\includegraphics[width=1.0\linewidth]{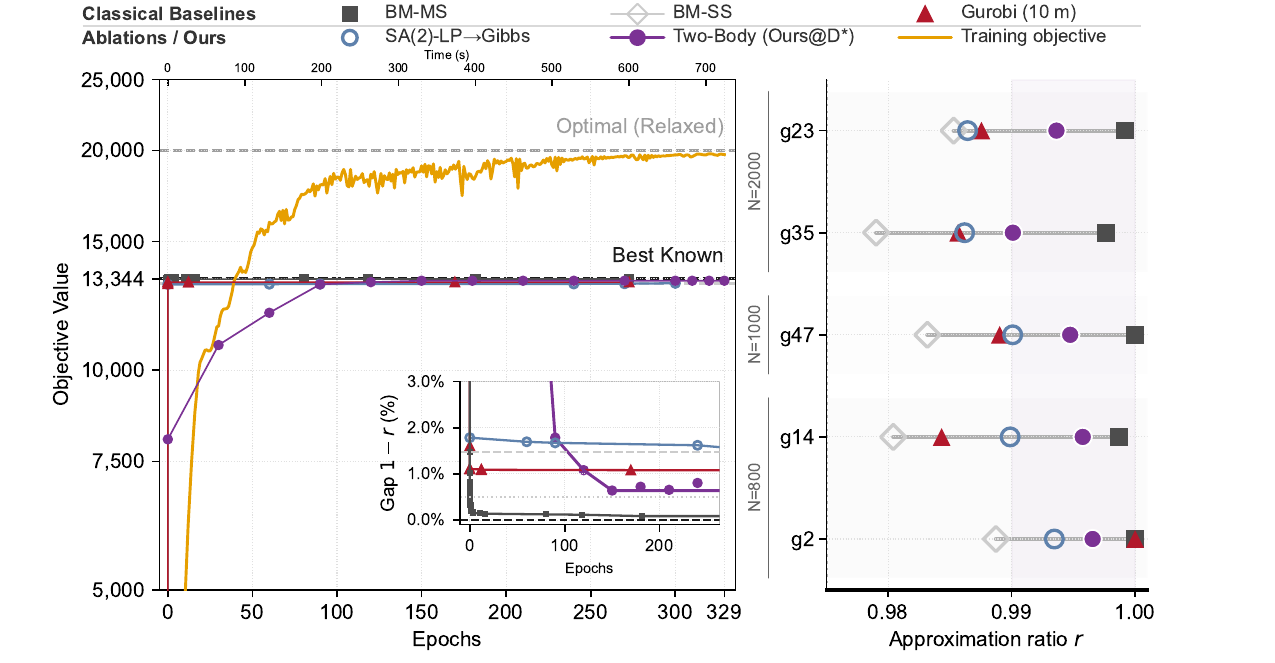}
\caption{
\textbf{Training dynamics and incumbent ratios on \textsc{Gset}.}
\emph{Left:}~Representative instance \textsc{G23} ($N{=}2000$).
Orange: training objective $E_{\mathrm{QUBO}}(\tilde\mu,\tilde\nu)$
(plotted as $-E_{\mathrm{QUBO}}$ so higher values correspond to better cuts).
Purple dots: decoded samples at sampling epochs.
Purple line: incumbent curve (running best).
Gray: BM-MS/BM-SS. Red: Gurobi (10\,min). Blue outline: SA(2)$\!\to$\,Gibbs.
Horizontal dashed lines: relaxation optimum from the $\mathrm{SA}(2)$--LP and the
best-known cut obtained by the Breakout Local Search (BLS) heuristic~\cite{benlic_breakout_2013}.
Shaded band: $\approx0.99$ region.
(Specs: $n_q{=}24$, $D{=}2$, 46~two-qubit gates.)
\emph{Right:}~Incumbent approximation ratios $r^\star$ across \textsc{Gset}
($N{=}800$--$2000$) at the median-optimal depth~$D^\ast$
($D^\ast{=}2$; \textsc{G35}:~$D^\ast{=}3$).
Gray: BM. Red: Gurobi. Blue outline: SA(2)$\!\to$\,Gibbs.
Purple: our solver, Two-Body~(@$D^\ast$).
Vertical rules group instances by~$N$.
Shaded band: constant-depth plateau.
All results use the frozen policy
($\rho{=}0.5$, $\lambda_{\mathrm{KL}}\!:\!0.10{\rightarrow}0.30$);
only depth varies on \textsc{Gset}.
Gibbs decode only; no local search or classical post-processing is applied.}
\label{fig:gset_headliner}
\end{figure}

\paragraph{Structured benchmarks: \textsc{Gset}.}
The \textsc{GSET} suite was held out entirely from hyper-parameter tuning;
only circuit depth~$D$ is varied to test for any depth-modulated effects.
Figure~\ref{fig:gset_headliner} summarizes both training dynamics
and incumbent approximation ratios across the suite.
The left panel shows a representative run on \textsc{G23} ($N{=}2000$):
the training objective $E_{\mathrm{QUBO}}(\tilde\mu,\tilde\nu)$ (orange)
rises near-monotonically, decoded incumbents (purple) saturate within the first ~150~epochs,
and the best-so-far curve stabilizes well before the relaxed objective. This run corresponds to the \emph{median seed} among five random initializations for $N{=}2000$; ten seeds were used for the smaller $N{=}800$ and $N{=}1000$ cases,
chosen to illustrate typical behavior.
The right panel compiles the best incumbent ratios $r^\star$ across all instances at the median-optimal depth $D^\ast{=}2$
(\textsc{G35}: $D^\ast{=}3$),
defined as the depth where the median gap across seeds is smallest; the reported value is the best-of-seed at that depth, a robust measure that
avoids cherry-picking. All other hyperparameter and training settings are held fixed; only depth varies between runs.
The shaded band marks the low-depth performance ($r^\star{\approx}0.99$).
All results use the frozen policy
($\rho{=}0.5$, $\lambda_{\mathrm{KL}}\!:\!0.10{\rightarrow}0.30$);
only depth varies.
Gibbs decode only; no local search or classical polishing is applied.

\FloatBarrier

\paragraph{Solver behavior on Erd\H{o}s--R\'enyi random graphs.}\label{subsec:ER-results-dyanmics}

The hyper-parameter sweeps that defined the frozen policy for \textsc{GSET} also
reveal the internal interactions that govern solver behavior.
Erd\H{o}s--R\'enyi (ER) graphs provide a statistically neutral manifold for these
diagnostics: random yet well-characterized, with controllable degree
statistics and no structural bias.
We focus here on the interplay between the projection strength~$\rho$ of
the under-relaxed IPF step and the circuit depth~$D$, and how their
interaction shapes the optimization gap.

We illustrate these effects at $N{=}128$, which is large enough for nontrivial structure and depth-dependent behavior to emerge, yet small enough to allow exhaustive sweeps over depth $D$ and projection strength~$\rho$.
Even with two layers ($D{=}2$), the solver consistently achieves approximation ratios above $r{\approx}0.94$, shown here as the gap to optimal—corresponding to gaps below the classic PCP inapproximability bound of $1/17{\approx}0.059$—confirming that constant-depth circuits suffice to reach the practically relevant regime.
Beyond that point, performance gains are marginal; deeper circuits instead illuminate the solver’s dynamics.
The sweeps suggest that $\rho$ acts as a geometric regularizer whose influence is confined to the shallow-depth regime.
At small~$D$, increasing~$\rho$ accelerates convergence and reduces the gap by nearly an order of magnitude relative to the unprojected control.
However, once circuit expressivity becomes sufficient ($D{\approx}10$--$14$ for $N{=}128$), the projected and unprojected trajectories converge, indicating that explicit feasibility repair is no longer necessary.
Beyond this depth, the circuit appears to internalize pairwise consistency on its own—the $\mathrm{SA}(2)$ geometry is effectively embedded within the ansatz.

\begin{figure}[t]
    \centering
    \includegraphics[width=1.0\linewidth]{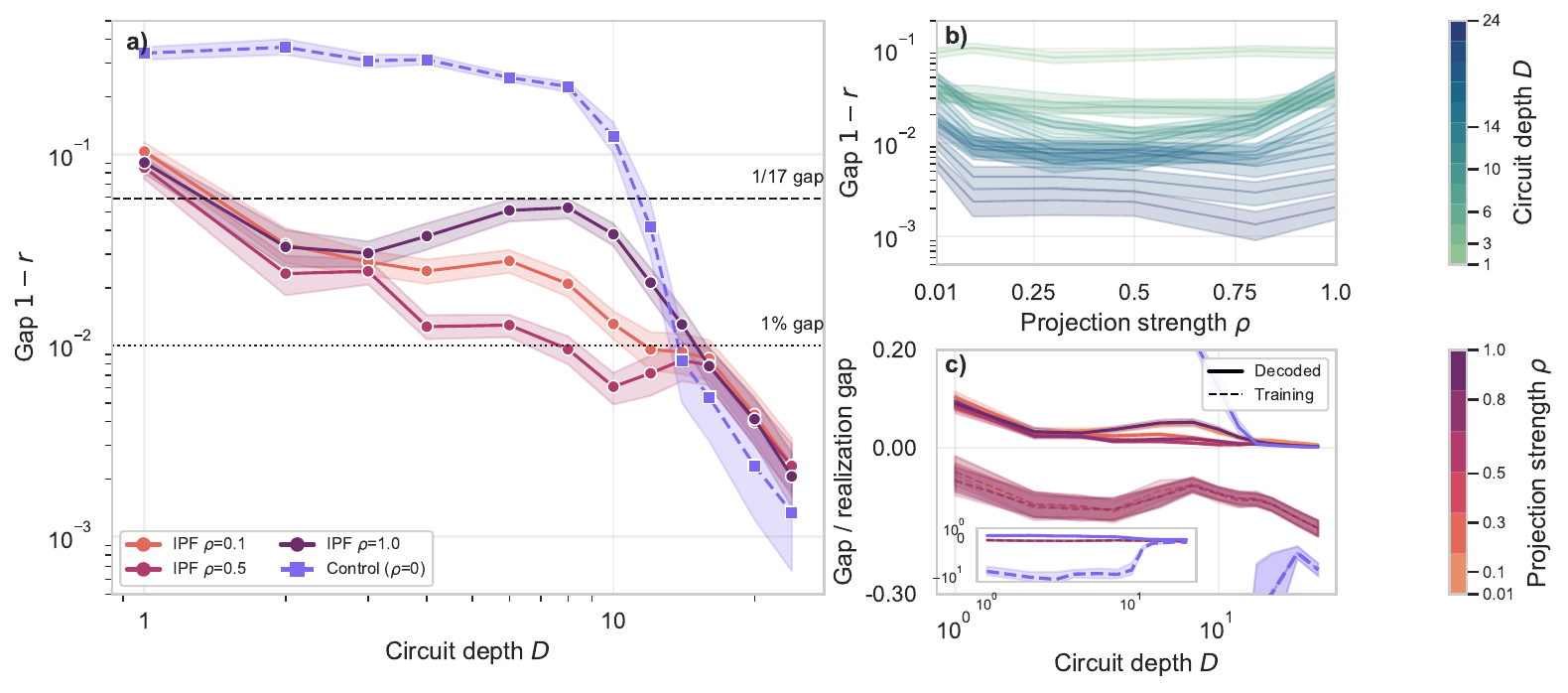}
    \caption{
    \textbf{Depth–accuracy and projection–strength effects on Erdos Renyi random graphs.}
    (a)~Gap $(1{-}r)$ versus depth~$D$ for representative $\rho$ values.
    Projected runs cross the PCP bound ($1/17{\approx}0.059$) at
    $D{\le}2$; the unprojected control catches up only at large~$D$.
    (b)~Gap versus~$\rho$, aggregated across depth bands, showing the
    characteristic U-shape with a broad optimum near~0.5.
    (c)~At $\rho{=}0.5$, decoded (solid) and training (dashed) objectives
    converge with depth, marking the transition to self-consistent
    geometry.
    }
    \label{fig:flagship_depth_vs_gap}
\end{figure}

Figure~\ref{fig:flagship_depth_vs_gap} illustrates these trends.
Panel~(a) shows the transition: all IPF variants rapidly fall below
the $1/17$ bound, while the unprojected control catches up only at large
depth.
Panel~(b) plots the characteristic U-shape in gap versus~$\rho$, with a
broad minimum near~0.5 that motivates the chosen frozen setting.
Panel~(c) reveals the tightening of the training and decoded objectives as
depth increases—the solver moves from a loose, exploratory phase to a
self-consistent one where repaired and raw moments coincide.
Together these diagnostics show that $\rho$-damped IPF governs the
low-expressivity regime, while deeper circuits naturally enter a
self-consistent plateau where feasibility is learned rather than imposed.

\paragraph{Scaling on Random Graphs} We also tested scaling from $N{=}32$ to $N{=}1024$ to verify that the observed low-depth behavior persists across system size.
The results are largely unremarkable: both the raw and running-best gaps
remain largely flat with~$N$, confirming that the same geometric regime governs
small and moderate graphs alike.
Because these curves offer no additional insight beyond confirming stability,
we defer the scaling plots to App.~\ref{app:ablation}
(Fig.~\ref{fig:er_depth_step_final_overlay}).
Scalability at larger~$N$ is already evident from the \textsc{Gset}
benchmarks, which extend the same low-depth behavior to instances up
to $N{=}2000$.

%% file: sections/09-discussion-outlook-epilogue.tex
\clearpage
\section{The Information Minimal Framework}

The surprise is not simply that the two-layer, real-amplitude quantum circuit performs well, but how clearly it succeeds once we separate out concerns of representation, regularization, and decoding.
By conventional standards—and our own initial expectations—such a minimal ansatz should have been insufficient: too shallow and constrained by real amplitudes to approach near-optimal performance. Indeed, without the IPF projection, the ansatz behaves exactly as theory would suggest, approximating a shallow random circuit and demonstrating limited expressivity consistent with approximate two-design behavior on random (Erdős–Rényi) graphs. Yet, once explicitly combined with IPF, the same minimalist circuit achieves approximation ratios above 0.99 on structured (\textsc{Gset}) graphs, rivaling far deeper and more sophisticated architectures—without intricate training routines, specialized decoding, or hidden classical heuristics.

Embedding local consistency and maximum-entropy decoding directly into the loss would overload the quantum structure, forcing it to learn representation, constraints, and decoding simultaneously—precisely the complexity minimalism seeks to avoid. But such an approach would overload the quantum structure, forcing it to simultaneously learn to encode representations, enforce constraints, and handle decoding—precisely the complexity minimalism seeks to avoid. Instead, each minimal element—quantum circuit, geometric projection, and classical decoder—occupies exactly its necessary role, no more. Complexity emerges not because we explicitly engineer it into any single element, but naturally from the interaction of simple, clearly defined parts.

\paragraph{Less is more.}
We deliberately stripped the quantum ansatz down to bare bones—two layers, real amplitudes, and hardware-efficient wiring—to clearly expose its limits. Without the projection and regularization, the circuit aligns with conventional depth–expressivity intuition, requiring greater depth ($D$) to approach useful performance, consistent with approximate 2-design behavior on random Erdős–Rényi graphs. Yet, with only a single $\rho$-damped IPF step to enforce local consistency, the performance curve flattens dramatically: at $D{=}2$–$3$, the method already reaches near-optimal approximation ratios (> 0.99) on structured \textsc{Gset} graphs up to $N{=}2000$, while additional depth provides little improvement or even diminishes performance. Here, depth stops signifying increased representational power ("reach"), and instead marks how quickly the quantum ansatz aligns with the underlying geometry. Each minimal component—the shallow circuit, the real amplitudes, the geometric projection—empirically captures the space allowed by the problem constraints.

\paragraph{More is different.}
This effectiveness does not arise from adding quantum ornamentation but from how minimal parts cooperate once their roles are clearly separated. 
The circuit learns an effective pairwise model—an empirical Hamiltonian $(\widehat{h},\widehat{J})$—that captures the low-dimensional moment structure of the problem; 
IPF projects this model onto the locally consistent body of $\mathrm{SA}(2)$; 
and Gibbs sampling then realizes an ensemble of bitstrings that inhabit this surface, the maximum-entropy expression of its geometry. 
All three stages are ``cost-blind,'' geometric, and probabilistic. 
On \textsc{Gset} benchmarks, two ablations illustrate this interplay: pairing a linear $\mathrm{SA}(2)$ relaxation with Gibbs , 
and sampling directly from the raw QUBO in Ising form (Table \ref{tab:gset_appendix_by_method}). 
Both perform worse—the latter markedly so, even at higher inverse temperatures—while the full pipeline consistently improves both the objective and the sampled cut quality.

The pattern suggests a thermodynamic picture. 
The circuit and IPF identify a point $(\mu^*,\nu^*)$ in moment space that minimizes the objective, 
and the MaxEnt decoding reparameterizes the local geometry around that point—effectively ``warming'' a region that was cold and rigid under the original QUBO Hamiltonian into one that is smooth and sampleable. 
This process does not modify the problem Hamiltonian but acts on the \textit{learned} $(\widehat{h},\widehat{J})$, 
the surrogate correlations inferred by the learned representation. 
The Gibbs sampler then populates this surface with bitstring configurations drawn from its consistent ensemble, 
linking optimization and sampling as facets of the same geometric process. 
\textit{Complexity, in this view, is not engineered but emergent—arising from the cooperation of simple, self-consistent parts that, together, form a richer whole.}

\paragraph{The informational-minimal geometry.}
Empirically, the pipeline defines an informationally minimal geometric framework in which each component occupies precisely its necessary role: the quantum circuit encodes pairwise correlations as pseudo-moments; the geometric projection explicitly ensures local consistency within the minimal SA(2) constraints; the Gibbs sampler explicitly realizes the encoded geometry as a maximum-entropy distribution. None of these elements alone captures the observed adequacy; performance emerges from their carefully defined interplay. Together, they isolate exactly what minimal statistics—pairwise correlations bounded by local consistency—are the minimal information needed to represent and learn the optimization problem. In other words, empirical adequacy is not about embedding more quantum complexity, but about clearly revealing the minimal geometry that the problem demands.

\paragraph{From qubit-efficiency to information-efficiency.}
This minimal geometry already demonstrates a clear principle of information efficiency.
All results presented here were obtained in the \emph{analytic statevector regime}, 
deliberately chosen to isolate geometric clarity from sampling complexity and hardware noise. 
The exponential compression afforded by our approach—representing $N$ variables using only $O(\log N)$ qubits—implies a fundamental exchange: 
exponential savings in representation may demand proportional increases in sampling complexity. 
Concretely, each measurement now yields only $O(\log N)$ bits of information rather than $N$, implying that more shots may be required to recover reliable statistics for optimization. 
Quantifying this sampling cost under finite-shot and realistic hardware conditions is therefore an essential direction for future work. 
Even in analytic form, however, the present results reveal a broader conceptual point: 
rather than competing solely on qubit-efficiency, we can now begin to chart a more fundamental landscape of \emph{information efficiency}—how representation, resources, and information interplay.

\section{Charting the Information-Efficient Frontier}

If such a minimal representation already achieves empirical adequacy, the next question is how this adequacy scales under limited resources—how efficiency trades off among representational capacity (width), sampling complexity (shots), and optimization effort (trainability).
So far we have isolated geometry in an analytic state-vector regime; the broader principle of information efficiency remains to be explored in finite-shot and hardware-realistic settings, as we chart the landscape of representation, resources, and learning.

\paragraph{Three natural extensions...}  This geometric logic naturally extends along three axes. \textit{Width} --  raising the Sherali–Adams level from SA(2) to SA(k) tightens QUBO bounds and reaches higher-order HUBOs while keeping the register count modest—roughly $(k(\lceil \log_2 N \rceil+1))$ qubits—and generalizing both IPF and Gibbs to higher-order marginals and coupling tensors. \textit{Shots} --analyzing finite-sampling complexity and hardware baselines becomes essential—under compression, each measurement carries less information but admits massive parallelism, and multiple log-width circuits could be run concurrently on present devices. \textit{Trainability}—exponential compression turns linear expectations into ratios of Born probabilities, introducing curvature that couples expressivity and learnability; measuring and regularizing this curvature through adaptive updates or smoother circuit families offers a route to stable training on weighted and frustrated graphs, and can be further benchmarked on the broader, multi-problem \emph{Decathlon} suite~\cite{koch_quantum_2025}, which spans increasingly difficult optimization tasks. More broadly, testing classical refinements—Mirror descent \cite{beck_mirror_2003}, Sinkhorn transport \cite{cuturi_sinkhorn_2013}, or parallel tempering \cite{newman_monte_2010}—will reveal how much they can elevate an otherwise minimal quantum representation. Together these axes define how the same minimal geometry can scale—testing how far adequacy persists before curvature, not order or sampling, becomes the limiting factor.

\paragraph{...And a fourth, to resist.}
It is tempting to imagine adding “more quantum’’—by layering non-local observables or multi-basis encodings in the hope of greater space efficiency. Doing so might conveniently shift the label from “quantum-inspired’’ to “quantum’’ in the usual taxonomy. 
Yet the target we optimize remains the same pairwise-moment body $\mathrm{SA}(2)$.
The space spanned by first- and second-order statistics $(\mu,\nu)$ has dimension $M=\mathcal{O}(N^2)$, and our logarithmic-width circuit already represents this body within a real Hilbert space of size 
$2^{2\lceil\log_2 N\rceil+2}\!\approx\!4N^2\!=\!\mathcal{O}(N^2)$.
Multi-basis schemes may shave constant factors by “stuffing’’ the same pairwise information into more elaborate Hilbert spaces, but this is bookkeeping, not geometry.  
Until the feasible set itself bends—until the polytope gives way to a spectrahedron—no amount of complex amplitude introduces new structure; it remains the same story written in fancier ink.  
The true frontier therefore lies not in ornamentation but in curvature: when the geometry itself begins to bend, and quantumness becomes structure rather than style.

\paragraph{The information-efficient frontier.}  
Taken together, width, shots, and trainability span a three-axis trade-off mirroring space, parallel time, and sequential time. Width—and its dynamic partner, circuit depth—sets the representational capacity and qubit resources of the model; shots determine how much information can be gathered in parallel between updates; and trainability measures how efficiently those updates navigate the curvature that compression introduces into the optimization landscape.

Every variational method spends this information budget differently: QAOA invests heavily in qubits and depth to keep its objective linear and gradients straightforward, while our compressed scheme explicitly trades qubits for a non-linear objective defined by ratios of Born probabilities.  
Such non-linearity remains benign in the analytic limit but could become a principal cost in finite-shot, non-simulatable conditions, where curvature and sampling noise directly interact. Improving along one dimension inevitably imposes stress along another—stronger compression saves width but increases curvature; better gradient estimates can navigate curvature but demand more shots.  
This surface of explicit, interlocking costs defines an \emph{information-efficient frontier}: a space–time geometry in which each compressed encoding, ansatz, and optimization method is clearly one experiment in how quantum and classical resources explicitly exchange.  
Charting this frontier—especially at points where curvature becomes irreducible and simulation is no longer straightforward—is the collective empirical and theoretical task ahead.

\paragraph{Scope and regime.}
Within the taxonomy introduced earlier, our framework is a single, end-to-end \textit{compression primitive} of encoding the full QUBO objective in logarithmic width, without decomposition or coarsening.
Large-scale decomposition methods—such as multilevel QAOA~\cite{maciejewski_multilevel_2024, bach_solving_2025}—address scale by splitting instances into subproblems before refinement, while compression primitives like Pauli-Correlation Encoding (PCE)~\cite{sciorilli_towards_2025} reduce dimensionality through shared Pauli correlators.
We belong to this latter class and operate in the same benchmark regime—unweighted Max-Cut on \textsc{GSET}—used across these studies for large-scale claims. Our contribution should therefore be read not as a universal solver but as an existence proof: in this saturated, pairwise regime a real, constant-depth, information-minimal circuit already achieves empirical adequacy.
Weighted and frustrated QUBOs, including Maximum Independent Set, plausibly demand richer geometric structure—curvature, interference, or non-commutation—and those remain open frontiers for future work. In that sense, our results tighten what future demonstrations can claim: any genuine quantum improvement must now operate where the flat, pairwise geometry fails and the feasible body bends toward a spectrahedron.

\medskip
The question that follows is sharper still: what kind of geometry truly demands quantumness rather than merely admits it?

\section*{Epilogue --- Beyond the Polytope, the Spectrahedron}

The geometry of quantum mechanics is not linear but curved—a spectrahedron, not a polytope.
Beyond the limits of $\mathrm{SA}(2)$, and more generally $\mathrm{SA}(k)$ at any finite level, lies a curved body defined by positive-semidefinite correlations,
\begin{align*}
      X_{ij}=\langle v_i , v_j \rangle,\qquad X \succeq 0,\quad X_{ii}=1.
\end{align*}
Many optimization families live more naturally in this geometry, where curvature encodes coherence and the onset of non-commutation. Curvature, in this view, is not incidental -- it is the operational signature of genuinely quantum structure.

Developing compressed, PSD-aware analogues of our two-body framework—perhaps with low-rank correlation models or multi-basis ($X/Y/Z$) measurements— becomes the natural continuation of the same geometric logic.
Extending the compression paradigm into this curved domain, first for two-body and ultimately for $k$-body PSD relationships, would mark the transition from a contestable quantum-inspired primitive to a definitively quantum one, completing the geometric progression outlined.

\paragraph{In Sum.} This work delineates a plateau and a path.
For a representative class of unweighted QUBOs, a two-body, polyhedral, real-amplitude geometry captures what matters in practice.
The task now is to chart where this empirical adequacy fails and to identify the first problem families whose tight, tractable relaxations are intrinsically spectrahedral—where the geometry itself bends and quantumness becomes necessary.
Beyond that point, quantum structure is not an addition but a continuation: the natural curvature of a deeper convex order.
Progress will come from letting geometry set the agenda -- and from introducing complexity only when the problem’s own structure demands it.

% This three-dimensional frontier describes trade-offs among spatial, parallel, and sequential resources,
% but the framework our algorithm occupies on this surface is itself real—its amplitudes, statistics, and updates all live in $\mathbb{R}$.  
% By contrast, models such as QAOA inhabit curved, complex-valued spaces where interference and phase provide additional degrees of freedom.  
% Is there, then, a \emph{quantumness axis} to this same frontier—one that measures not how much quantum structure we can \emph{add}, but how much the problem itself \emph{requires}?  

% \section{Toward a Theory of Information flow in Variational Learning under Compressed Encodings}

% Zoom back into the original intent of VQE. How to interpret that result under the compression literature.

% \paragraph{From Optimization to Representation Learning }

% The information-efficient frontier, first outlined here, provides not just a way to optimize, but a new lens through which to see variational quantum algorithms:

%  Optimization as representation learning, width as compression, and decoding complexity as the intrinsic cost of information retrieval.

%% file: sections/99-appendix.tex
\begin{appendix}
% \section{Finite-sample estimators and smoothing}\label{app:finite-shots}
% Let \(C_{ijab}\) be counts over \(M\) shots and \(\hat p_{ijab}=C_{ijab}/M\). The plug-in estimators \(\hat\mu,\hat\nu\) are defined by Eqs.~\eqref{eq:mu}–\eqref{eq:nu} with \(p\) replaced by \(\hat p\). We use a small \(\varepsilon>0\) in denominators for stability. A simple \emph{addressing-prior} can reduce variance when \(\mathcal{O}_i\) (or \(\mathcal{O}_{\{i,j\}}\)) is rare: add one virtual observation of value \(1/2\) to each denominator (and \(1/2\) to the corresponding numerator), which keeps the estimator unbiased under symmetric ignorance and prevents division by zero. We found this practical in large-\(N\) regimes where many addressed pairs might be unobserved at low shot budgets.%
\input{sections/appendix/a1_quantum_moment_encoding}

\input{sections/appendix/a2_sa2}

\input{sections/appendix/a3_ipf}
\input{sections/appendix/a4_ablations}

\input{sections/appendix/a5_gset_tables}
\end{appendix}

%% file: sections/appendix/a1_quantum_moment_encoding.tex
\section{Quantum Form of Moment Encoding}\label{app:quantum-version-of-moments}
The main text uses the geometric/probabilistic language because the flat,
pairwise regime requires only classical consistency conditions.
Nevertheless, all quantities can be written in standard quantum form:
expectations $\mathrm{Tr}[\rho_\theta\,\Pi]$ of joint projective measurement elements on
$\mathcal H_I\!\otimes\!\mathcal H_A\!\otimes\!\mathcal H_J\!\otimes\mathcal H_B$.
We include this formulation not as a translation exercise, but as a
foundation for future work where the feasible set becomes spectrahedral
and the relevant variables—positive semidefinite operators—may not
commute.

Let $\rho_\theta = |\psi(\theta)\rangle\!\langle\psi(\theta)|$ be the circuit state.
Define address projectors
$\Pi^I_i = |i\rangle\!\langle i|_I$, $\Pi^J_j = |j\rangle\!\langle j|_J$,
and value projectors
$\Pi^A_a = |a\rangle\!\langle a|_A$, $\Pi^B_b = |b\rangle\!\langle b|_B$.
The joint event projector for
$(I=i,A=a,J=j,B=b)$ is
\[
\Pi_{iajb}
   = \Pi^I_i\!\otimes\!\Pi^A_a\!\otimes\!\Pi^J_j\!\otimes\!\Pi^B_b,
   \qquad
   p_\theta(i,a,j,b)
   = \mathrm{Tr}[\rho_\theta\,\Pi_{iajb}].
\]

\paragraph{Conditioning and “bit = 1” effects.}
For $i\neq j$, define the address-conditioning and “bit = 1” effects as
\[
\begin{aligned}
P^{(1)}_i
   &= \sum_{j\neq i}\!\Big(
      \Pi^I_i\!\otimes\!\mathbb I^A\!\otimes\!\Pi^J_j\!\otimes\!\mathbb I^B
     +\Pi^I_j\!\otimes\!\mathbb I^A\!\otimes\!\Pi^J_i\!\otimes\!\mathbb I^B
     \Big), \\[2pt]
E^{(1)}_i
   &= \sum_{j\neq i}\!\Big(
      \Pi^I_i\!\otimes\!\Pi^A_1\!\otimes\!\Pi^J_j\!\otimes\!\mathbb I^B
     +\Pi^I_j\!\otimes\!\mathbb I^A\!\otimes\!\Pi^J_i\!\otimes\!\Pi^B_1
     \Big), \\[4pt]
P^{(2)}_{ij}
   &= \Pi^I_i\!\otimes\!\mathbb I^A\!\otimes\!\Pi^J_j\!\otimes\!\mathbb I^B
     +\Pi^I_j\!\otimes\!\mathbb I^A\!\otimes\!\Pi^J_i\!\otimes\!\mathbb I^B, \\[2pt]
E^{(2)}_{ij}
   &= \Pi^I_i\!\otimes\!\Pi^A_1\!\otimes\!\Pi^J_j\!\otimes\!\Pi^B_1
     +\Pi^I_j\!\otimes\!\Pi^A_1\!\otimes\!\Pi^J_i\!\otimes\!\Pi^B_1.
\end{aligned}
\]

\paragraph{Address-conditioned pseudo-moments.}
The conditional Born expectations are
\[
\hat\mu_i
   = \frac{\mathrm{Tr}[\rho_\theta\,E^{(1)}_i]}
           {\mathrm{Tr}[\rho_\theta\,P^{(1)}_i]},
   \qquad
\hat\nu_{ij}
   = \frac{\mathrm{Tr}[\rho_\theta\,E^{(2)}_{ij}]}
           {\mathrm{Tr}[\rho_\theta\,P^{(2)}_{ij}]}.
\]
Each $\hat\mu_i,\hat\nu_{ij}\in[0,1]$, but the full set
$\{\hat\mu,\hat\nu\}$ need not be pairwise-feasible;
different pairs are conditioned on distinct address events,
hence the need for a Bregman/IPF projection to obtain a feasible
moment vector $(\tilde\mu,\tilde\nu)\in\mathrm{SA}(2)$.

\paragraph{Local $2\times2$ tables.}
Define the local joint-probability table for the unordered pair $\{i,j\}$:
\[
\widehat{P}^{\{i,j\}}_{ab}
  = \frac{\mathrm{Tr}[\rho_\theta\,M^{\{i,j\}}_{ab}]}
           {\mathrm{Tr}[\rho_\theta\,P^{(2)}_{ij}]},
  \qquad
  M^{\{i,j\}}_{ab}
  = \Pi^I_i\!\otimes\!\Pi^A_a\!\otimes\!\Pi^J_j\!\otimes\!\Pi^B_b
   +\Pi^I_j\!\otimes\!\Pi^A_b\!\otimes\!\Pi^J_i\!\otimes\!\Pi^B_a.
\]
The ancilla swap in the second term ensures that the first index
always corresponds to $x_i$ and the second to $x_j$.
Non-negativity of all four entries of
$\widehat{P}^{\{i,j\}}$ reproduces the Boole--Fr\'echet bounds.

\paragraph{Energy expectation.}
The quadratic energy expectation is then
\[
E_Q(\hat\mu,\hat\nu)
   = \sum_i Q_{ii}\,\hat\mu_i
     + 2\sum_{i<j} Q_{ij}\,\hat\nu_{ij},
\]
identical to the geometric form used in the main text.

The operator form presented here reproduces exactly the quantities
defined in the main text; the two notations are equivalent in content.
We include it not as a translation exercise but as groundwork:
when the geometry extends beyond the flat polytope to the curved,
positive-semidefinite spectrahedron, operator-valued variables and
trace-based constraints will become the natural geometry of the problem.

%% file: sections/appendix/a2_sa2.tex
\section{Equivalence of Sherali–Adams level-2 (SA(2)) and Boole–Fréchet bounds}
\label{app:sa2}

\paragraph{Setup.} Let $X=(X_1,\dots,X_N)\in\{0,1\}^N$ be binary variables.
We write
\(
\mu_i := \mathbb{E}[X_i]\in[0,1]
\)
and
\(
\nu_{ij} := \mathbb{E}[X_iX_j]\in[0,1]
\)
for $i\neq j$. (For completeness, $X_i^2=X_i$ implies $\nu_{ii}=\mu_i$, though in the main text we store only off-diagonals and set $\nu_{ii}$ by convention.)

\subsection{Sherali–Adams hierarchy (formal definition)}
Consider a $0$–$1$ feasibility set
\(
\mathcal{F}=\{x\in\{0,1\}^N: A x \le b\}.
\)
The level-$k$ Sherali–Adams \cite{sherali_hierarchy_1994} relaxation $\mathrm{SA}(k)$ is obtained by
(i) taking each original inequality $a^\top x\le b$, together with the box constraints
$0\le x_i\le 1$,
(ii) multiplying it by all monomials
\(
\prod_{i\in U}x_i \prod_{j\in V}(1-x_j)
\)
with $U,V\subseteq[N]$, $U\cap V=\emptyset$, and $|U|+|V|\le k-1$,
and (iii) \emph{linearizing} each resulting polynomial by introducing a variable for every monomial of degree at most $k$ and replacing products by those variables. The projection of the resulting lifted polytope onto the degree-$\le 2$ coordinates yields $\mathrm{SA}(2)$.

For the special case with only box constraints (no $Ax\!\le\!b$ beyond $0\le x_i\le 1$), the degree-$2$ coordinate set is precisely
\(
\{x_i,\ y_{ij}=x_ix_j:\ 1\le i\le N,\ 1\le j\le N\},
\)
and $\mathrm{SA}(2)$ is defined by all linear consequences obtained from box constraints after the above lifting and linearization.

\subsection{Specialization to binary pairs: Boole–Fréchet bounds}
Applying the level‑2 lifting to the box constraints $x_i\ge 0$ and $1-x_i\ge 0$ by multiplying with $x_j$ and $(1-x_j)$ ($j\ne i$) yields the following four linear inequalities linking the degree‑1 variables $x_i$ and the degree‑2 variables $y_{ij}$:
\begin{align}
x_i x_j &\ge 0 &&\Longrightarrow && y_{ij}\ \ge\ 0, \label{eq:bf-1}\\
x_i(1-x_j) &\ge 0 &&\Longrightarrow && y_{ij}\ \le\ x_i, \label{eq:bf-2}\\
(1-x_i)x_j &\ge 0 &&\Longrightarrow && y_{ij}\ \le\ x_j, \label{eq:bf-3}\\
(1-x_i)(1-x_j) &\ge 0 &&\Longrightarrow && y_{ij}\ \ge\ x_i+x_j-1. \label{eq:bf-4}
\end{align}
Together with $0\le x_i\le 1$, Eqs.~\eqref{eq:bf-1}–\eqref{eq:bf-4} are exactly the \emph{Boole–Fréchet} bounds for the bilinear terms $y_{ij}=x_ix_j$.
Under the identification $x_i\leftrightarrow \mu_i$ and $y_{ij}\leftrightarrow \nu_{ij}$, these become
\begin{equation}
\max\{0,\ \mu_i+\mu_j-1\}\ \le\ \nu_{ij}\ \le\ \min\{\mu_i,\ \mu_j\},
\qquad 0\le \mu_i\le 1,\quad i\ne j.
\label{eq:bf-mu-nu}
\end{equation}

\subsection{Equivalence by construction: SA(2) $\iff$ Boole–Fréchet}
\begin{lemma}[Pairwise table reconstruction]{Lemma (pairwise table reconstruction).}
Given $(\mu_i,\mu_j,\nu_{ij})$ with $0\le \mu_i,\mu_j\le 1$, define a $2\times 2$ table
\[
P_{ij} \;=\;
\begin{array}{c|cc}
 & x_j=0 & x_j=1\\\hline
x_i=0 & 1-\mu_i-\mu_j+\nu_{ij} & \ \mu_j-\nu_{ij}\\[2pt]
x_i=1 & \mu_i-\nu_{ij} & \ \nu_{ij}
\end{array}\, .
\]
Then $P_{ij}$ is a valid joint distribution (nonnegative entries summing to $1$) with
$\sum_{x_j}P_{ij}(1,x_j)=\mu_i$ and $\sum_{x_i}P_{ij}(x_i,1)=\mu_j$
\emph{if and only if} the Boole–Fréchet bounds Eq.~\eqref{eq:bf-mu-nu} hold.

\smallskip
\noindent\emph{Proof.}
($\Rightarrow$) Nonnegativity of each entry implies
$\nu_{ij}\ge 0$,
$\mu_i-\nu_{ij}\ge 0$,
$\mu_j-\nu_{ij}\ge 0$, and
$1-\mu_i-\mu_j+\nu_{ij}\ge 0$, which are exactly \eqref{eq:bf-mu-nu}.
($\Leftarrow$) If \eqref{eq:bf-mu-nu} holds, each entry of $P_{ij}$ is nonnegative and the row/column sums match $\mu_i,\mu_j$ by construction. \qed
\end{lemma}
\medskip
\begin{proposition}{Proposition (SA(2) $\iff$ Boole–Fréchet for box‑only constraints).}
For binary variables with only box constraints, the $\mathrm{SA}(2)$ relaxation projected to $(x_i,y_{ij})$ coincides with the Boole–Fréchet polytope:
\[
\mathrm{SA}(2)\ =\ \Big\{(x,y)\ \Big|\ 0\le x_i\le 1,\ \max\{0,x_i+x_j-1\}\le y_{ij}\le \min\{x_i,x_j\}\ \ \forall i\ne j\Big\}.
\]

\smallskip
\noindent\emph{Proof sketch.}
($\subseteq$) By the level‑2 lifting of box constraints, any point in $\mathrm{SA}(2)$ satisfies Eqs.~\eqref{eq:bf-1}–\eqref{eq:bf-4}, hence Eq.~\eqref{eq:bf-mu-nu}.  
($\supseteq$) Conversely, if Eq.~\eqref{eq:bf-mu-nu} holds for all $i\ne j$ together with $0\le x_i\le 1$, then (by the lemma) each pair $(i,j)$ admits a valid $2\times2$ table with marginals $(x_i,x_j)$ and $y_{ij}$. These are precisely the degree‑$\le 2$ feasibility conditions generated by the level‑2 lifting of the box; no additional linear consequences at degree $\le 2$ remain. \qed
\end{proposition}
\subsection{Remarks}
\begin{itemize}
\item \textbf{Local (pairwise) polytope.} The feasible set defined Eq.~\eqref{eq:bf-mu-nu} for all edges, together with the row/column sum identities, is the pairwise local marginal polytope. It is exact on trees and a relaxation on loopy graphs.
\item \textbf{Diagonal convention.} Algebraically $y_{ii}=x_i$ in SA(2); in our moment parameterization we store only $i\ne j$ and set $\nu_{ii}=0$ by convention for the two‑body matrix used in the energy (the diagonal contribution is carried by $\mu_i$).
\item \textbf{Relation to QUBO.} With $x_i\leftrightarrow \mu_i$ and $y_{ij}\leftrightarrow \nu_{ij}$, $\mathrm{SA}(2)$ provides the tightest linear constraints expressible solely at the one‑ and two‑body level, which is exactly the granularity at which $\mathbb{E}[x^\top Q x]$ depends on the distribution.
\end{itemize}

%% file: sections/appendix/a3_ipf.tex
\section{KL/IPF Projection} \label{app:ipf}

\paragraph{Theoretical background.}
We use “Iterative Proportional Fitting (IPF)” in a generalized sense—as alternating KL/Bregman projections between convex constraint sets.
The term is kept for simplicity: it denotes the same mechanics as classical IPF but without fixed marginals.
On convex polyhedral regions such as $\mathrm{SA}(2)$, these alternating KL projections converge to the unique I-projection, the point of minimal Bregman (KL-type) divergence to the original iterate~\cite{heinz_h_bauschke_method_1997}.
Because the pseudo-moments $(\widehat\mu,\widehat\nu)$ are expectations rather than normalized probabilities, the update acts in the Bregman sense under the negative-entropy potential, not as a probabilistic KL between joint tables.
In effect, it treats $(\mu,\nu)$ as coordinates in a convex information geometry and projects them under the same divergence that defines classical IPF.
The $\rho$-damped variant used during training approximates this fixed-point solution in one or a few steps while preserving differentiability. 

\paragraph{Notation.}
For notational clarity, we write $(\mu,\nu)$ for the projected (feasible) moments,
corresponding to $(\tilde{\mu},\tilde{\nu})$ in the main text.
Throughout this appendix, $(\widehat{\mu},\widehat{\nu})$ denote the raw pseudo-moments
and $(\mu,\nu)$ their $\mathrm{SA}(2)$-projected counterparts obtained by the IPF update.
\emph{Subsequent appendices revert to the notation of the main text, where
$(\widehat{\mu},\widehat{\nu})$ are pseudo-moments and
$(\tilde{\mu},\tilde{\nu})$ denote their repaired versions.}

\paragraph{Problem statement (I‑projection).}
Given raw pseudo‑marginals $(\widehat{\mu},\widehat{\nu})$ in $[0,1]\times[0,1]$, that is not necessarily feasible for Eq.\eqref{eq:boole-frechet}, we define the KL I‑projection onto $\mathrm{SA}(2)$ as
\begin{equation}
(\mu^\star,\nu^\star)
\;\in\;
\arg\min_{(\mu,\nu)\ \text{s.t.}\ \eqref{eq:boole-frechet},\ \nu=\nu^\top,\ \nu_{ii}=0}
\Big\{
D_{\mathrm{KL}}(\mu\|\widehat{\mu}) + D_{\mathrm{KL}}(\nu\|\widehat{\nu})
\Big\},
\label{eq:kl-proj}
\end{equation}
where $D_{\mathrm{KL}}$ is applied elementwise and summed:
\begin{align}
    D_{\mathrm{KL}}(\mu\|\widehat{\mu})
=\sum_i \mu_i\log\frac{\mu_i}{\widehat{\mu}_i}-\mu_i+\widehat{\mu}_i
\end{align}\label{eqn:bregman_distance}

and similarly for $\nu$ (e.g.\ over $i<j$).
The feasible set in Eq.\eqref{eq:kl-proj} is a polytope; the objective is strictly convex on its relative interior, so the solution is unique whenever it exists.

\paragraph{$\rho$‑damped KL/IPF iteration (inequality‑only form).}
A light‑weight projector that solves \eqref{eq:kl-proj} by alternating Bregman projections uses two ``snaps’’ per iteration. With damping parameter $\rho\in(0,1]$, the iterate update is
\[
\nu \leftarrow (1-\rho)\,\nu + \rho\,\nu^{\mathrm{snap}},\qquad
\mu \leftarrow (1-\rho)\,\mu + \rho\,\mu^{\mathrm{snap}}.
\]
Setting $\rho=1$ recovers classical IPF/Dykstra‑style steps; $\rho<1$ yields a smoother projector useful in gradient‑based training.
The procedure is $O(|E|)$ per pass for sparse graphs (or $O(N^2)$ dense), uses only elementwise clips, and is differentiable almost everywhere (piecewise‑linear with well‑behaved subgradients at the clip points). A practical implementation follows this scheme verbatim.

\paragraph{Stopping and diagnostics.}
A simple early‑exit rule is $\max_i|\Delta\mu_i|<\text{tol}$ (with tol $>0$ small).
An \emph{edgewise KL gap} can be reported to quantify progress:
\[
\mathrm{KL\_gap} \;=\; 
\sum_i D_{\mathrm{KL}}(\mu_i\|\widehat{\mu}_i)
\;+\;
\sum_{i<j} D_{\mathrm{KL}}(\nu_{ij}\|\widehat{\nu}_{ij}),
\]
evaluated at the current $(\mu,\nu)$.
It decreases monotonically along exact IPF steps and serves as a smooth penalty in training.

\paragraph{Algorithmic summary.}
\begin{algorithm}[h]
\caption{KL/IPF projection onto SA(2)}
\label{alg:ipf}
\begin{algorithmic}[1]
\Require Raw pseudo‑marginals $(\widehat{\mu},\widehat{\nu})$, damping $\rho\in(0,1]$, iterations $T$, tolerance $\text{tol}>0$.
\State Initialize $\mu\leftarrow\operatorname{clip}(\widehat{\mu},\varepsilon,1-\varepsilon)$, $\nu\leftarrow\operatorname{clip}(\widehat{\nu},\varepsilon,1-\varepsilon)$; set $\nu\leftarrow\tfrac12(\nu+\nu^\top)$, $\nu_{ii}\leftarrow0$.
\For{$t=1,\dots,T$}
  \State \emph{($\nu$‑box snap)} For all $i\neq j$:
  \State \quad Compute $L_{ij}\leftarrow \max\{0,\mu_i+\mu_j-1\}$, 
         $U_{ij}\leftarrow \min\{\mu_i,\mu_j\}$
  \State \quad Set $\nu^{\mathrm{snap}}_{ij}\leftarrow\operatorname{clip}(\nu_{ij},L_{ij},U_{ij})$
  \State \quad Symmetrize: $\nu^{\mathrm{snap}}\leftarrow\tfrac12(\nu^{\mathrm{snap}}+(\nu^{\mathrm{snap}})^\top)$, 
         $\nu^{\mathrm{snap}}_{ii}\leftarrow0$
  \State \hspace{1.3em} Damped update: $\nu\leftarrow(1-\rho)\nu+\rho\,\nu^{\mathrm{snap}}$.
  \State \emph{($\mu$‑consistency snap)} For each $i$:
  \State \quad Compute $L_i\leftarrow\max_{j\neq i}\nu_{ij}$, 
         $U_i\leftarrow\min_{j\neq i}(1+\nu_{ij}-\mu_j)$
  \State \quad Set $\mu^{\mathrm{snap}}_i\leftarrow\operatorname{clip}(\mu_i,L_i,U_i)$ 
         then clamp to $[0,1]$
  \State \hspace{1.3em} Damped update: $\mu\leftarrow(1-\rho)\mu+\rho\,\mu^{\mathrm{snap}}$.
  \State If $\max_i|\Delta\mu_i|<\text{tol}$: \textbf{break}.
\EndFor
\State \Return $(\mu,\nu)$ and (optionally) $\mathrm{KL\_gap}$.
\end{algorithmic}
\end{algorithm}

\paragraph{Remarks.}
(i) The projection never materializes $2\times2$ tables; all operations are pointwise clips dictated by \eqref{eq:boole-frechet}.  
(ii) Symmetry and zero diagonal are enforced after each snap.  
(iii) On polyhedral constraints, alternating KL projections converge to the I‑projection; damping offers a smooth trade‑off between feasibility and gradient flow useful in end‑to‑end learning.

\section{Benign Initialization and Gradient Flow Through the $\rho$-damped Repair}
\label{app:ipf-followup}

Two practical questions motivate a single, under-relaxed repair per step:
(i) are we ever \emph{starting} from a pathological corner such as $(\varepsilon,\varepsilon,1-\varepsilon)$, and
(ii) if $(\hat\mu,\hat\nu)$ lies \emph{outside} $\mathrm{SA}(2)$, do gradients still flow through the repair map?
This appendix gives short answers: initialization is benign, and a $\rho$-damped snap preserves (and re-routes) gradient signal.

\paragraph{Notation and setting.}
Hats $(\hat\mu,\hat\nu)$ denote the pseudo-moments from Sec.\ref{subsec:parameters-to-pseudomoments};
tildes $(\tilde\mu,\tilde\nu)$ are the moments \emph{after one} $\rho$-damped IPF snap (§\ref{sec:sa2-ipf}), used in the loss
$E_{\mathrm{QUBO}}(\tilde\mu,\tilde\nu)=\sum_i Q_{ii}\tilde\mu_i+2\sum_{i<j}Q_{ij}\tilde\nu_{ij}$.
The repair consists of a $\nu$-box followed by a $\mu$-consistency snap:
\[
\boxed{
\begin{aligned}
\tilde\nu_{ij} &= (1-\rho)\,\hat\nu_{ij}\;+\;\rho\,\Pi_{[L_{ij}(\hat\mu),\,U_{ij}(\hat\mu)]}\!\big(\hat\nu_{ij}\big),\\[2pt]
\tilde\mu_i     &= (1-\rho)\,\hat\mu_i\;+\;\rho\,\Pi_{[L_i(\tilde\nu),\,U_i(\hat\mu,\tilde\nu)]}\!\big(\hat\mu_i\big),
\end{aligned}}
\qquad
\begin{aligned}
&L_{ij}(\hat\mu)=\max\{0,\hat\mu_i+\hat\mu_j-1\},\;\; U_{ij}(\hat\mu)=\min\{\hat\mu_i,\hat\mu_j\},\\
&L_i(\tilde\nu)=\max_{j\neq i} \tilde\nu_{ij},\;\; U_i(\hat\mu,\tilde\nu)=\min_{j\neq i}\bigl(1+\tilde\nu_{ij}-\hat\mu_j\bigr).
\end{aligned}
\]
We re-enforce symmetry and set $\tilde\nu_{ii}=0$ after damping.

\subsection{Benign Start: Uniform Hadamard Initialization and a Robust Tail Model}
\paragraph{Hadamard start.}
With an initial Hadamard layer (§\ref{sec:moment-encoding}), the measurement distribution is uniform on $\{0,1\}^N$:
\[
\hat\mu_i=\tfrac12,\qquad \hat\nu_{ij}=\tfrac14,\qquad
L_{ij}(\hat\mu)=0,\quad U_{ij}(\hat\mu)=\tfrac12.
\]
Thus every edge starts \emph{strictly inside} the Boole–Fréchet box; gross pathologies are impossible at $D{=}0$.
\paragraph{Dirichlet tails (when randomness is present).}
Beyond the exact H-init point, we model the initialization outcome probabilities as draws from a symmetric Dirichlet ensemble—the distribution over outcome probabilities induced by Haar-like (2-design) circuit amplitudes.
This approximation captures the statistics of generic deep-circuit initialization and bounds the probability of extreme moment values.
The resulting scatter ratios are Beta-distributed, yielding exponentially small tails for configurations such as $(\varepsilon,\varepsilon,1-\varepsilon)$:
\[
\hat\mu_i=\frac{p(A_i)}{p(D_i)}\sim\mathrm{Beta}(a_i,b_i),\qquad
\hat\nu_{ij}=\frac{p(A_{ij})}{p(D_{ij})}\sim\mathrm{Beta}(c_{ij},d_{ij}),
\]
with shape parameters determined by the effective ensemble size.
Standard Beta tails give, for any $\varepsilon\in(0,1)$,
\[
\Pr[\hat\mu_i\le\varepsilon]\le \frac{\varepsilon^{\,a_i}}{a_i B(a_i,b_i)},\qquad
\Pr[\hat\nu_{ij}\ge 1-\varepsilon]\le \frac{\varepsilon^{\,d_{ij}}}{d_{ij} B(c_{ij},d_{ij})}.
\]
In our encoding $a_i,b_i=\Theta(N)$, so triples like $(\varepsilon,\varepsilon,1-\varepsilon)$ are vanishingly unlikely as $N$ grows.
\emph{Takeaway:} at or near random initialization, heavy repair is not expected.

\subsection{Backpropagation Through the Under–Relaxed IPF}
We record only the Jacobians relevant for the objective function here; subgradients are taken at clip/max/min boundaries.

\paragraph{Coordinate-wise derivatives.}
For the $\nu$-box at fixed $\hat\mu$,
\[
\frac{\partial \tilde\nu_{ij}}{\partial \hat\nu_{ij}}
=(1-\rho)+\rho\,\mathbf{1}\!\big\{L_{ij}(\hat\mu)<\hat\nu_{ij}<U_{ij}(\hat\mu)\big\},
\]
so on any \emph{saturated} edge (clip chose a bound) this equals $(1-\rho)$ (and is $0$ when $\rho{=}1$).
Bound-dependence creates cross-terms:
\[
\frac{\partial \tilde\nu_{ij}}{\partial \hat\mu_i}=
\begin{cases}
\rho, & U_{ij}=\hat\mu_i \;\text{and}\; \hat\nu_{ij}>U_{ij},\\
\rho, & L_{ij}=\hat\mu_i+\hat\mu_j-1>0 \;\text{and}\; \hat\nu_{ij}<L_{ij},\\
0, & \text{otherwise,}
\end{cases}
\qquad
\frac{\partial \tilde\nu_{ij}}{\partial \hat\mu_j}\ \text{is symmetric.}
\]
For the singleton snap over $[L_i(\tilde\nu),U_i(\hat\mu,\tilde\nu)]$,
\[
\frac{\partial \tilde\mu_i}{\partial \hat\mu_i}
=(1-\rho)+\rho\,\mathbf{1}\{L_i<\hat\mu_i<U_i\},\quad
\frac{\partial \tilde\mu_i}{\partial \tilde\nu_{ij}}\in
\begin{cases}
\{\rho\}, & j\in\arg\max_{k\neq i}\tilde\nu_{ik}\ \text{and}\ \tilde\mu_i=L_i,\\
\{-\rho\},& j\in\arg\min_{k\neq i}(1+\tilde\nu_{ik}-\hat\mu_k)\ \text{and}\ \tilde\mu_i=U_i,\\
\{0\}, & \text{otherwise.}
\end{cases}
\]

\paragraph{What this implies for learning.}
Let $g_i=\partial E/\partial \tilde\mu_i$ and $g_{ij}=\partial E/\partial \tilde\nu_{ij}$.
Then:
\begin{itemize}
\item \textbf{Direct signal never vanishes for $0{<}\rho{<}1$.}
On any saturated edge $(i,j)$, the direct energy-path derivative through $\hat\nu_{ij}$ scales exactly by $(1-\rho)$.
\item \textbf{Signal re-routed when bounds are active.}
The cross-terms above feed a $\rho$-weighted gradient into the singletons whenever an active bound depends on $\hat\mu$—precisely where feasibility binds.
\item \textbf{Why under–relaxation helps.}
Full clipping ($\rho{=}1$) zeroes the direct $\nu$-path on saturated edges; under–relaxation preserves a fixed fraction of that signal and adds a stabilizing singleton path.
\end{itemize}

\subsection{One–step contraction of violations (used in the $\rho$-sweep)}
Write edgewise violations $v^+_{ij}=(\hat\nu_{ij}-U_{ij})_+$ and $v^-_{ij}=(L_{ij}-\hat\nu_{ij})_+$.
After the $\nu$-box,
\[
v^{+\,\mathrm{new}}_{ij}=(1-\rho)\,v^+_{ij},\qquad v^{-\,\mathrm{new}}_{ij}=(1-\rho)\,v^-_{ij},
\]
so the total violation mass contracts by a factor of $(1-\rho)$ in one snap.
This explains the near-linear $\rho$ trend in Fig.~\ref{fig:flagship_depth_vs_gap}(b).
\medskip

\noindent\emph{Bottom line.} The uniform Hadamard-initialization puts us at the center of $\mathrm{SA}(2)$; the Dirichlet model shows pathologies are exceedingly unlikely nearby.
When training drifts outside $\mathrm{SA}(2)$, a single under-relaxed repair both \emph{preserves gradients} and \emph{prevents exploits} of pathological corners—sufficient for the dynamics we report, without a full convergence theory that we defer to future work.

%% file: sections/appendix/a4_ablations.tex
\section{Hyperparameter Calibration, Ablations, and Scaling Studies}
\label{sec:ablation}\label{app:ablation}

\input{tables/er_sweep_hyperparams}

Systematic sweeps on Erdős–Rényi (ER) graphs were used to identify stable training
regimes for projection strength~$\rho$, KL–penalty schedules, and learning–rate policies.
Unless otherwise noted, each configuration was evaluated across
$N\!\in\!\{32,64,128\}$, ten random seeds per expected degree $\alpha$, and multiple depths. Each run used a decoding budget proportional to the instance size,
with $10\,N\log_2N$ Gibbs samples per evaluation step, ensuring consistent sampling effort across $N$. For the extended learning–rate and scaling ablations
($N{=}32$–$1024$), we used five seeds per configuration to balance coverage and runtime.

\paragraph{KL and projection sweeps.}
We compared constant KL penalties ($0.1$) with linear ramps targeting
$\{0.3,1.0\}$ and observed no consistent improvement or degradation across settings.
The penalty factor had only a weak influence in this regime, so for simplicity we adopt a mild KL ramp $0.1{\rightarrow}0.3$ as a representative
setting for all subsequent experiments.

\subsection{Extended Depth–Projection Strength Analysis}
\label{app:rho-depth-facets}

\begin{figure}[h]
    \centering
    \includegraphics[width=\linewidth]{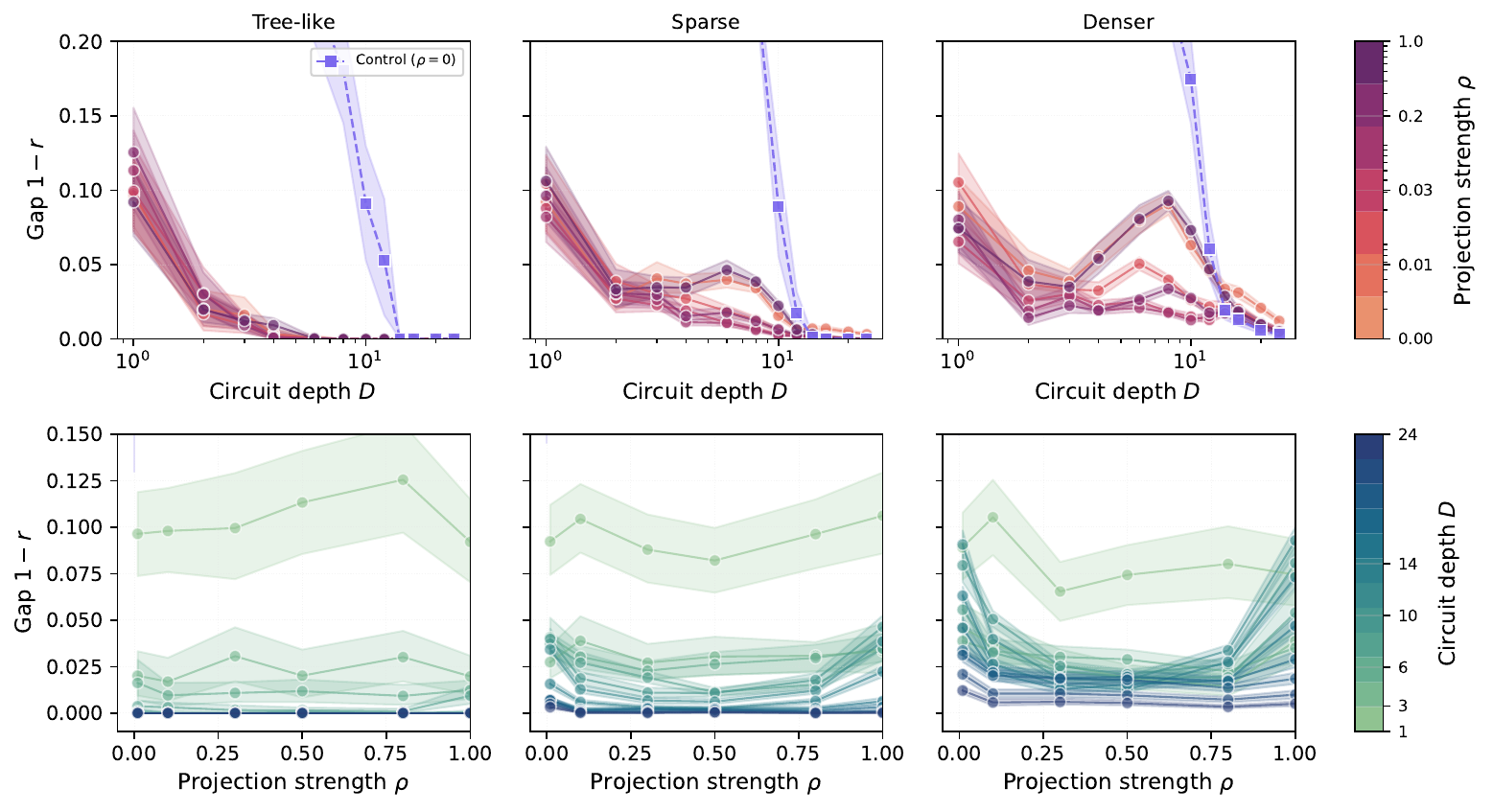}
    \caption{\textbf{Depth–projection trade-offs across graph densities (\(N{=}128\)).}
    Each column corresponds to one density family (tree-like, sparse, denser; 
    \(\alpha{=}1,\ 4\!-\!5,\ 8\!-\!12\)).
    Top: gap \(1{-}r\) versus circuit depth \(D\) for projection strengths 
    \(\rho\!\in[0,1]\); 
    bottom: gap versus projection strength for fixed depths. 
    Each curve aggregates five random seeds per configuration. 
    Trends mirror those in Fig.~\ref{fig:flagship_depth_vs_gap}: 
    under-relaxed IPF (\(\rho\!\approx\!0.3\!-\!0.5\)) consistently yields the smallest gaps, 
    with a characteristic U-shaped dependence on \(\rho\). 
    Higher-degree graphs display a sharper “knee’’ near \(\rho{=}0.5\) and slightly 
    noisier convergence, but the same qualitative pattern—diminishing sensitivity 
    at larger depths once the ansatz saturates. 
    These dynamics guided the choice of \(\rho{=}0.5\) for dense instances such as GSET.}
    \label{fig:appendix-rho-depth-facets}
\end{figure}

Figure~\ref{fig:appendix-rho-depth-facets} provides a more granular view of the 
projection–depth dynamics discussed in the main text (Fig.~\ref{fig:flagship_depth_vs_gap}). 
We separate the three graph families (\(\alpha{\in} \{1\},\ \{4,5\},\ \{8,12\}\) at fixed size \(N{=}128\),
showing how the optimization gap \(1{-}r\) evolves jointly with circuit depth \(D\) and
projection strength \(\rho\).
Across all densities, the pattern is consistent: shallow circuits exhibit a pronounced 
U-shape in $\rho$ whose minimum stabilizes around \(0.3\!-\!0.5\), 
while deeper circuits flatten the curve as the ansatz saturates and the IPF correction becomes redundant. 
Higher-degree graphs show slightly looser convergence and a sharper “knee’’ near 
\(\rho{\approx}0.5\), indicating that stronger regularization improves stability in dense regimes.
This behavior motivated our empirical default of \(\rho{=}0.5\) for the GSET suite, 
where graph densities are high.

\subsection{Training and Penalty Schedule}

All experiments share the same functional form for the KL-penalty ramp and
the learning-rate schedule.  Both are defined piecewise over the normalized epoch fraction~$t/T \in [0,1]$.

\paragraph{KL-penalty ramp.}
The penalty weight $\lambda_{\mathrm{KL}}(t)$ increases linearly between fixed
fractions of total training:
\begin{align}
\lambda_{\mathrm{KL}}(t)
&=
\begin{cases}
\lambda_{\text{start}}, &
t/T < f_{\text{start}}, \\[4pt]
\lambda_{\text{start}}
  + \displaystyle
    \frac{\lambda_{\text{end}} - \lambda_{\text{start}}}
         {f_{\text{end}} - f_{\text{start}}}
    \bigl(\tfrac{t}{T}-f_{\text{start}}\bigr), &
f_{\text{start}} \le t/T < f_{\text{end}}, \\[8pt]
\lambda_{\text{end}}, &
t/T \ge f_{\text{end}}.
\end{cases}
\end{align}
In all reported runs we use
$\lambda_{\text{start}}{=}0.10$,
$\lambda_{\text{end}}{=}0.30$,
$f_{\text{start}}{=}0.15$,
and $f_{\text{end}}{=}0.85$.
This ramp prevents the feasibility term from dominating early
optimization while enforcing consistency later.

\paragraph{Learning-rate schedule.}
The learning rate follows a \texttt{warmup-hold-exponential} profile:
\begin{align}
\eta(t)
&=
\begin{cases}
\eta_{\text{start}}
  + (\eta_{\text{peak}} - \eta_{\text{start}})
    \tfrac{t/T}{f_{\text{warm}}}, &
t/T < f_{\text{warm}}, \\[6pt]
\eta_{\text{peak}}, &
f_{\text{warm}} \le t/T < f_{\text{warm}} + f_{\text{hold}}, \\[6pt]
\eta_{\text{peak}}
  \exp\!\Bigl[-\kappa
  \bigl(\tfrac{t/T - f_{\text{warm}} - f_{\text{hold}}}
             {1 - f_{\text{warm}} - f_{\text{hold}}}\bigr)\Bigr], &
t/T \ge f_{\text{warm}} + f_{\text{hold}},
\end{cases}
\end{align}
where the decay constant $\kappa$ is chosen so that
$\eta(T)=\eta_{\text{end}}$.
Default parameters are
$\eta_{\text{start}}{=}0.03$,
$\eta_{\text{peak}}{=}0.10$,
$\eta_{\text{end}}{=}0.01$,
$f_{\text{warm}}{=}0.10$,
$f_{\text{hold}}{=}0.40$.
This schedule balances exploration and late-phase convergence,
and empirically yields faster stabilization of the training loss
relative to constant or purely exponential policies.

\begin{figure}[!ht]
  \centering
  \includegraphics[width=\linewidth]{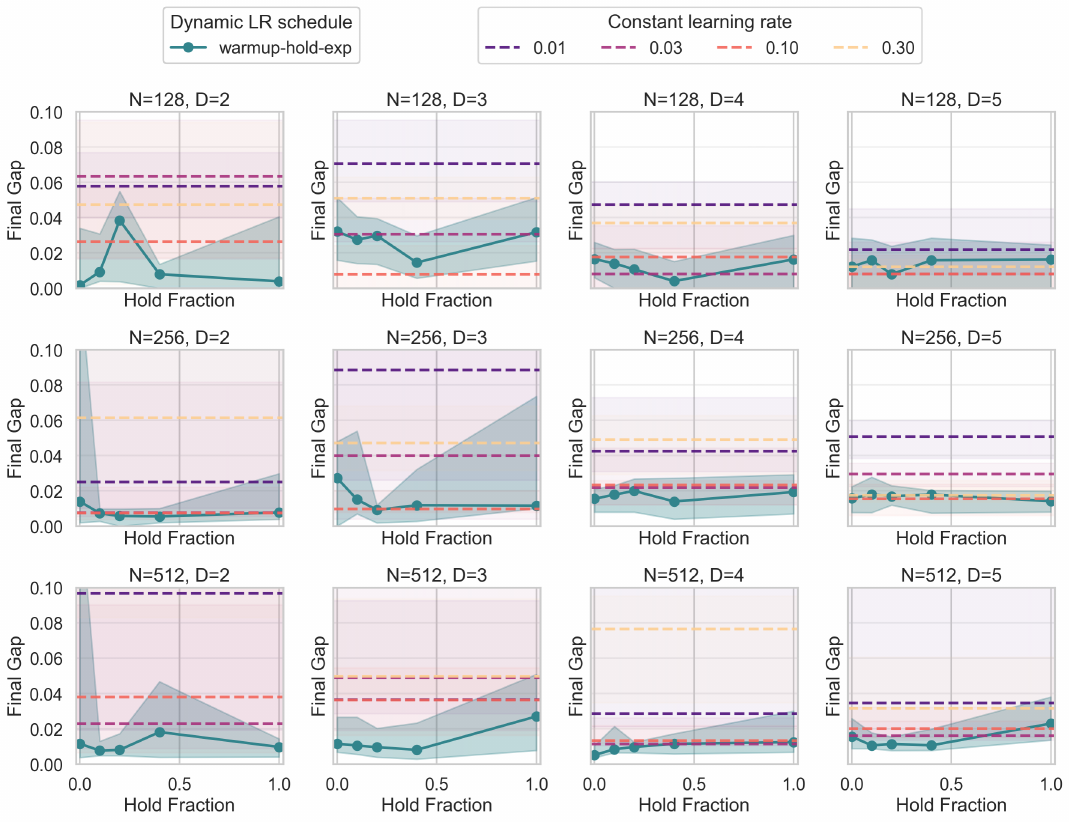}
  \caption{Final gap versus hold fraction for the warmup–hold–exponential
  schedule (solid teal) compared with constant learning rates (dashed).
  Each shaded band shows $\pm1$~s.d.\ over seeds.
  Dynamic schedules consistently yield lower final gaps across
  $N\!\in\!\{128,256,512\}$ and depths $D{=}2$–$5$.}
  \label{fig:lr_hold_ablation}
\end{figure}

\paragraph{Learning-rate sweep.}
Building on this configuration, we fixed $\rho{=}0.5$ and compared constant learning
rates to a dynamic \texttt{warmup-hold-exponential} schedule.  The dynamic schedule begins at
$\eta_{\mathrm{start}}{=}0.03$, rises linearly to $\eta_{\mathrm{peak}}{=}0.10$
over the first $10\%$ of epochs, holds for a fraction $f_h$ of the remaining epochs, then decays
exponentially to $\eta_{\mathrm{end}}{=}0.01$.
Hold fractions $f_h\!\in\!\{0.0,0.1,0.2,0.4,1.0\}$ were tested across
$N\!\in\!\{128,256,512\}$ and depths $D{=}2$–$5$.
Figure~\ref{fig:lr_hold_ablation} shows that constant learning rates
(\emph{dashed}) typically underperform, whereas the warmup–hold–exp policy consistently
reduces final gaps and improves late-phase stability.

\paragraph{Why this schedule and why 0.4?}
Training dynamics in variational circuits are non-convex
\cite{mcclean_theory_2016} and benefit from an initial exploratory
high-learning-rate phase, yet excessively large rates can destabilize
convergence.
Linear warm-up and decay are standard in machine learning \cite{goyal_accurate_2018}; adding a hold allows precise control
of the exploration window without altering the total epoch budget.
The warm-up phase also stabilizes Adam’s internal statistics and places the
optimizer in a well-conditioned region before high-LR exploration, while the
exponential decay ensures smooth convergence.
Notably, we found a peak learning rate of
$\eta_{\mathrm{peak}}{=}0.10$ to yield faster and more stable convergence
than the default Adam setting ($\eta{=}0.01$).

Figure~\ref{fig:lr_hold_ablation} shows no single hold fraction that dominates
across all $(N,D)$.
We therefore adopt $f_h{=}0.40$ as a modest, robust default:
it lies within~2\% of the per-block best on average, avoids the
instabilities of no-decay runs ($f_h{=}1.0$), and provides a generous
exploration window that scales well as circuit size grows.
Tables~\ref{tab:lr_warmup_vs_const_rho05}–\ref{tab:lr_hold_fraction_robust_rho05}
quantify this robustness: $f_h{=}0.4$ achieves the highest win-rate and
coverage within~2\%–5\% of the best constant-LR blocks, though overall
differences are small.
We standardize on the
\texttt{warmup$\rightarrow$hold(40\%)$\rightarrow$exp} schedule at
$\rho{=}0.5$, reducing hyperparameter search while maintaining performance
and stability.

\input{tables/appendix_lr_ablation_summary}
\input{sections/appendix/a4b_lr_ablation_tables}

% =========================================================
% Runtime scaling and sampling overhead
% =========================================================
\subsection{Runtime Scaling and Sampling Overhead.}
\begin{figure}[h]
  \centering
  \includegraphics[width=0.5\linewidth]{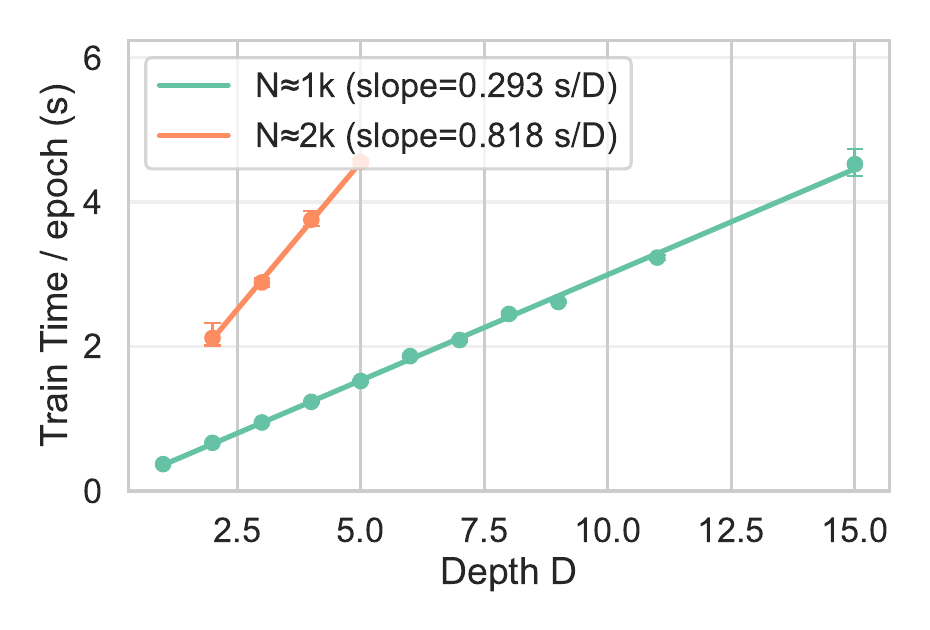}
  \caption{Average training time per epoch versus depth~$D$ for the
  GSET workloads used in the headline results.
  Points show the mean across seeds for the winning configuration and error
  bars indicate bootstrap 95\% intervals.  The fitted lines (legend) report
  the slope in seconds per extra layer, confirming linear $O(D)$ scaling.}
  \label{fig:train_time_depth}
\end{figure}

We also record wall-clock timings for the GSET benchmarks to verify that the
training policy remains lightweight. Figure~\ref{fig:train_time_depth}
plots the average time per epoch against depth~$D$ for the two size
families considered ($N\approx10^3$ and $N\approx2\times10^3$).
Each point aggregates the depth--$D$ runs that participate in the main results,
and the best-fit lines highlight the near-linear relationship with slope
approximately proportional to~$N$.  This behavior is expected from the
$O(N^2)$ cost of evaluating pairwise QUBO terms and the linear parameter growth in depth
of the hardware-efficient ansatz.

Taken together, these measurements demonstrate that the proposed schedule is
computationally light: training scales linearly with depth and problem size,
and decoding overhead is negligible compared to optimisation time.

\paragraph{Erdős--Rényi scaling experiment}

To verify that the low-depth regime observed in the main text persists with
problem size, we conducted an additional sweep over
$N{=}\{32,64,128,256,512,1024\}$.
All runs used the same calibrated learning-rate and KL-penalty schedules as
the ER experiments in Sec.~\ref{subsec:ER-results-dyanmics}
(\texttt{warmup--hold--exponential} schedule with
$f_h{=}0.40$, $\eta_{\mathrm{peak}}{=}0.10$,
and $\rho{=}0.5$).
Epoch counts were scaled linearly with~$N$
(120--300 epochs) to ensure comparable optimization effort.
At each depth we recorded both the instantaneous mean gap
and the running-best (incumbent) gap. The configuration matches the frozen training policy used throughout the paper.

\begin{figure}[h]
    \centering
    \includegraphics[width=0.8\linewidth]{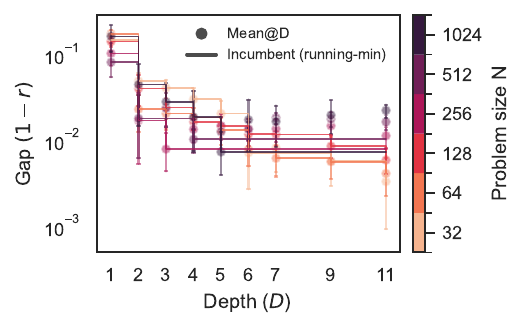}
    \caption{\textbf{Scaling of ER performance with problem size.}
    Running-best (solid) and raw mean (points with error bars) gap versus
    depth~$D$ for $N{=}32$--$1024$.
    The low-depth plateau identified in the main text persists uniformly with
    system size, indicating that the same geometric regime governs both small
    and moderate graphs.
    }
    \label{fig:er_depth_step_final_overlay}
\end{figure}

Figure~\ref{fig:er_depth_step_final_overlay} summarizes the results.
The running-best gap remains essentially constant across system sizes,
confirming that the same constant-depth plateau governs the solver’s
behavior from $N{=}32$ to $N{=}1024$.

Overall, these calibration and scaling studies validate the frozen policy used
throughout the main experiments:
a mild $\rho$-damped projection, linear KL ramp to~0.3, and the
\texttt{warmup$\!\rightarrow$hold(40\%)$\!\rightarrow$exp} learning-rate
schedule provide stable, transferable training dynamics across graph sizes
and densities.

%% file: tables/er_sweep_hyperparams.tex
\begin{table}[h]
\caption{Hyperparameter grid for the systematic ER sweep.}
\label{tab:er_sweep_hyperparams}
\begin{tabular}{ll}
\toprule
Parameter & Values \\
\midrule
$N$ & 32, 64, 128, 256, 512, 1024 \\
Expected degree $\alpha$ & 1, 4, 5, 8, 12 \\
Depth $D$ & 1, 2, 3, 4, 6, 8, 10, 12, 14, 16, 20, 24, 27, 30 \\
Projection $\rho$ & 0.00, 0.01, 0.1, 0.3, 0.5, 0.7, 0.8, 1.0 \\
KL target $\lambda_{\mathrm{end}}$ & 0.1, 0.3, 1.0 \\
\bottomrule
\end{tabular}
\end{table}

%% file: tables/appendix_lr_ablation_summary.tex
\begin{table}
\centering
\caption{Adam LR ablation summary on ER instances. Warmup--hold--exp (peak=0.10) beats the best constant LR baseline across all reported sizes, while hold fraction $0.40$ stays within $0.011$ of the block-wise optimum.}
\label{tab:lr_ablation_summary}
\begin{tabular}{rccc}
\toprule
$N$ & Warmup/Const Ratio & Improvement & Hold 0.40 Gap \\
\midrule
128 & 0.50 & 50\% & 0.0009 \\
256 & 0.33 & 67\% & 0.0107 \\
512 & 0.50 & 50\% & 0.0058 \\
\bottomrule
\end{tabular}
\end{table}

%% file: sections/appendix/a4b_lr_ablation_tables.tex
% Appendix: LR ablation summary tables (@ rho=0.5)
% This file is generated based on CSVs under experiments/lr_ablation_er/tables/

\begin{table}[h]
  \centering
  \small
  \caption{warmup--hold--exp vs best constant LR (final-gap ratio and win-rate), $\rho=0.5$. Ratio $<1$ favors warmup--hold--exp. Blocks are paired by $(N, p, D, \text{seed})$.}
  \label{tab:lr_warmup_vs_const_rho05}
  \begin{tabular}{lccc}
    \hline
    Group & Median ratio & Win-rate & Blocks \\
    \hline
    ALL   & 0.333 & 0.628 & 180 \\
    N=128 & 0.458 & 0.417 &  60 \\
    N=256 & 0.333 & 0.650 &  60 \\
    N=512 & 0.310 & 0.817 &  60 \\
    \hline
  \end{tabular}
\end{table}

\begin{table}[h]
  \centering
  \small
  \caption{Hold-fraction robustness within warmup--hold--exp, $\rho=0.5$. Median difference to block-wise best (lower is better) and coverage within 2\%/5\% of best.}
  \label{tab:lr_hold_fraction_robust_rho05}
  \begin{tabular}{lccccr}
    \hline
    Hold frac & Median diff & Top-1 & Within 2\% & Within 5\% & Samples \\
    \hline
    0.0 & 0.00554 & 0.311 & 0.694 & 0.867 & 180 \\
    0.1 & 0.00769 & 0.256 & 0.739 & 0.911 & 180 \\
    0.2 & 0.00619 & 0.344 & 0.761 & 0.922 & 180 \\
    0.4 & \textbf{0.00503} & \textbf{0.389} & \textbf{0.817} & \textbf{0.922} & 180 \\
    1.0 & 0.00863 & 0.261 & 0.672 & 0.844 & 180 \\
    \hline
  \end{tabular}
\end{table}

%% file: sections/appendix/a5_gset_tables.tex
\section{Detailed \textsc{GSET} Results}
\label{app:gset_tables}

This section reports, for each GSET instance, the depth $D^*$ chosen by the best median across seeds for the proposed TwoBody+Gibbs solver, alongside classical baselines. We list the mean/median/best percent approximation ratio (higher is better), with the number of runs per method, and include model size (qubits $n_q$) and two-qubit gate counts only for our method.

We additionally include a direct QUBO$\!\to$Ising Hamiltonian baseline decoded by Gibbs
sampling.
Without temperature tuning this formulation mixes poorly—the chain remains
effectively ``cold''—so we apply a robust coupling rescaling,
\begin{align*}
r_c
\;=\;
\mathrm{quantile}_{0.95}\!
\Bigl(
  \bigl\{
    \textstyle\sum_{j\neq i}|J_{ij}|
  \bigr\}_{i=1}^N
\Bigr),
\qquad
J \leftarrow J/r_c,\quad
h \leftarrow h/r_c,
\end{align*}
to keep coupling magnitudes numerically stable.
Even with this adjustment, the Gibbs chain remains trapped in local modes,
yielding sub-optimal approximation ratios.
These results are included in
Table~\ref{tab:gset_appendix_by_method} for reference but omitted from the
main figure \ref{fig:gset_headliner} for visual clarity.

\input{tables/gset_appendix_by_method_table}

\begin{figure}[h]
    \centering
    \includegraphics[width=0.95\linewidth]{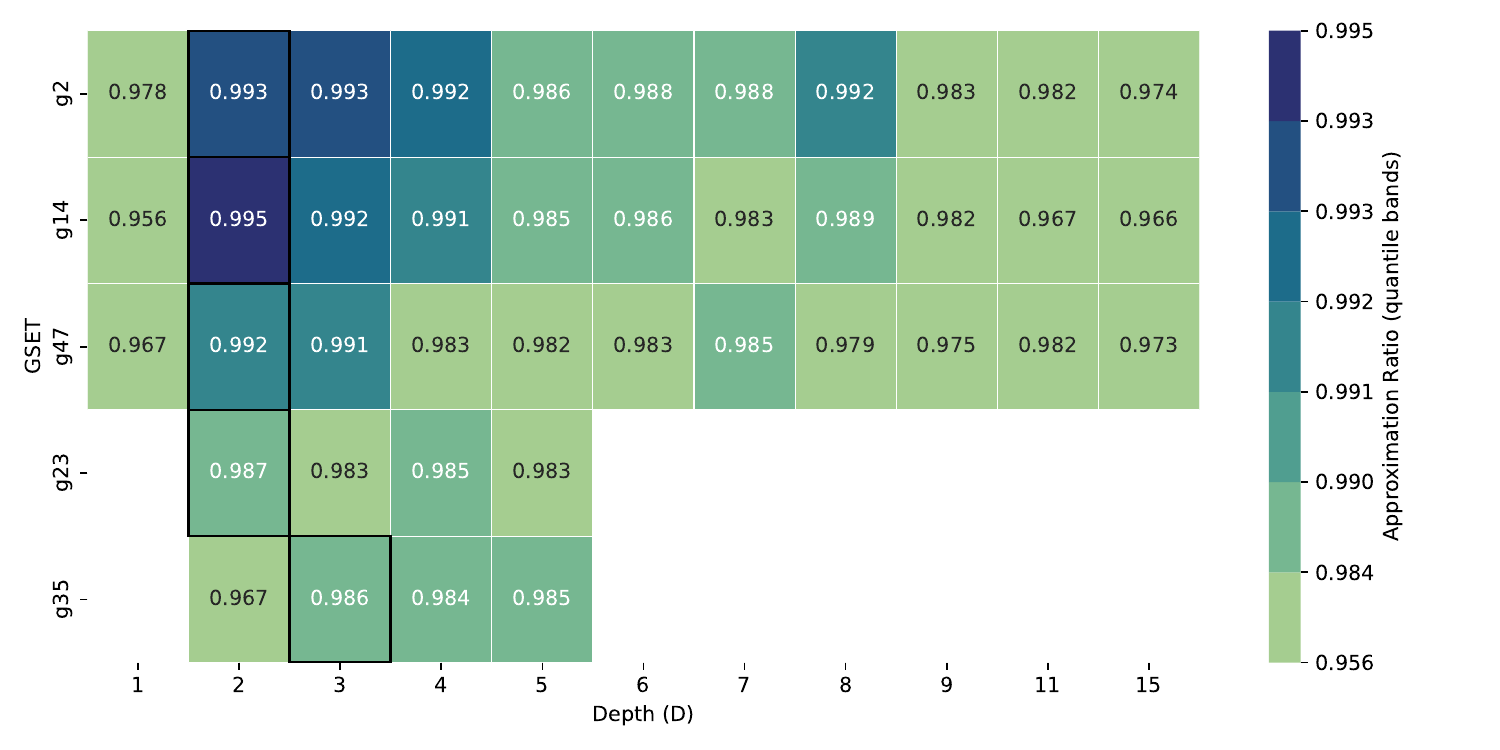}
    \caption{Approximation ratio for IPF across depths on GSET (median across seeds). Color scale emphasizes near 1 performance; outlined cells mark the best depth per problem instance based on the median approximation ratio.}
    \label{fig:appendix_ipf_fullperformance_heatmap}
\end{figure}

For completeness we also report the average Gibbs decode times and training durations at the depth $D^*$ used for each instance in the policy-parity baseline. Table~\ref{tab:gset_gibbs_time} lists the number of sweeps per chain (\mbox{$10$k} for $N\approx 10^3$, \mbox{$23$k} for $N=2000$), the selected depth, and the corresponding per-epoch and total training wall-clock time. Figure \ref{fig:appendix_ipf_fullperformance_heatmap} provides a summary of the GSET performance across the tested depth. For \texttt{g2, g14, g47} we conducted a more extensive sweep across depths, while the larger instances \texttt{g23, g35} we restricted to lower depths due to computational budget. 

\input{tables/gset_gibbs_time_summary}

Finally, for comparison, we also include the Table \ref{tab:gset_best_overall_raw_cuts} of the raw cuts value we obtained, the corresponding approximation ratios computed relative to the best-known 
solutions from the Breakout Local Search algorithm~\cite{benlic_breakout_2013}
\input{tables/gset_best_overall_table}

%% file: tables/gset_appendix_by_method_table.tex
% Auto-generated by render_appendix_by_method_table.py
\begin{table}[t]\centering
\caption{Approximation ratios at the depth $D^*$ (best median) for TwoBody+Gibbs and classical baselines on the GSET suite. Fig \ref{fig:gset_headliner} plots figures using the best approximation ratio for each instance. Values show approximation ratios (cut/Best-known). Method legend: GRB(10m)=Gurobi bound (10 min); QUBO$\to$Ising(robust)=Gibbs on scaled Ising Hamiltonian of the QUBO; SA2+Gibbs=SA(2)-LP+Gibbs; TB+Gibbs at $D^*$ (best median)}
\label{tab:gset_appendix_by_method}
\small
\begin{tabular}{llrrrrrr}\toprule
GSET & Method & $N$ & $D^*/runs$ & $n_q$/2Q@$D^*$ & \multicolumn{3}{c}{Approx.~ratio} \\
 \cmidrule(lr){6-8} 
 & & & & & mean & med & best \\
\midrule
 g2 & Gurobi (10m) & 800 & -/1.0 & 22/- & 1.000 & 1.000 & 1.000 \\
  & Burer-Monteiro & 800 & -/1.0 & 22/- & 1.000 & 1.000 & 1.000 \\
  & QUBO→Ising+Gibbs & 800 & -/1.0 & 22/- & 0.416 & 0.416 & 0.416 \\
  & SA(2)-LP+Gibbs & 800 & -/10.0 & 22/- & 0.993 & 0.993 & 0.993 \\
  & TwoBody+Gibbs (ours) & 800 & 2/10.0 & 22/42 & 0.976 & 0.993 & 0.997 \\
\midrule
 g14 & Gurobi (10m) & 800 & -/1.0 & 22/- & 0.984 & 0.984 & 0.984 \\
  & Burer-Monteiro & 800 & -/1.0 & 22/- & 0.999 & 0.999 & 0.999 \\
  & QUBO→Ising+Gibbs & 800 & -/1.0 & 22/- & 0.608 & 0.608 & 0.608 \\
  & SA(2)-LP+Gibbs & 800 & -/10.0 & 22/- & 0.988 & 0.988 & 0.990 \\
  & TwoBody+Gibbs (ours) & 800 & 2/10.0 & 22/42 & 0.992 & 0.995 & 0.996 \\
\midrule
 g47 & Gurobi (10m) & 1000 & -/1.0 & 22/- & 0.989 & 0.989 & 0.989 \\
  & Burer-Monteiro & 1000 & -/1.0 & 22/- & 1.000 & 1.000 & 1.000 \\
  & QUBO→Ising & 1000 & -/1.0 & 22/- & 0.388 & 0.388 & 0.388 \\
  & SA(2)-LP+Gibbs & 1000 & -/10.0 & 22/- & 0.987 & 0.987 & 0.990 \\
  & TwoBody+Gibbs (ours) & 1000 & 2/10.0 & 22/42 & 0.992 & 0.992 & 0.995 \\
\midrule
 g23 & Gurobi (10m) & 2000 & -/1.0 & 24/- & 0.988 & 0.988 & 0.988 \\
  & Burer-Monteiro & 2000 & -/1.0 & 24/- & 0.999 & 0.999 & 0.999 \\
  & QUBO→Ising+Gibbs & 2000 & -/1.0 & 24/- & 0.358 & 0.358 & 0.358 \\
  & SA(2)-LP+Gibbs & 2000 & -/5.0 & 24/- & 0.986 & 0.986 & 0.986 \\
  & TwoBody+Gibbs (ours) & 2000 & 2/5.0 & 24/46 & 0.978 & 0.987 & 0.994 \\
\midrule
 g35 & Gurobi (10m) & 2000 & -/1.0 & 24/- & 0.986 & 0.986 & 0.986 \\
  & Burer-Monteiro & 2000 & -/1.0 & 24/- & 0.998 & 0.998 & 0.998 \\
  & QUBO→Ising+Gibbs & 2000 & -/1.0 & 24/- & 0.525 & 0.525 & 0.525 \\
  & SA(2)-LP+Gibbs & 2000 & -/5.0 & 24/- & 0.986 & 0.986 & 0.986 \\
  & TwoBody+Gibbs (ours) & 2000 & 3/5.0 & 24/69 & 0.978 & 0.986 & 0.990 \\
\midrule
\bottomrule
\end{tabular}

\end{table}

%% file: tables/gset_gibbs_time_summary.tex
% Auto-generated by make_gset_gibbs_time_table.py
\begin{table}[h]\centering
\caption{Average Gibbs decode time and training time for the policy-parity baseline.}
\label{tab:gset_gibbs_time}
\small
\begin{tabular}{lrrrrrrrr}\toprule
GSET & $N$ & $|E|$ & Epochs & Sweeps & Depth* & Avg.~decode (s) & Avg.~train (s) & Train/epoch (s) \\
\midrule
 g14 & 800 & 4694 & 300 & 10000 & 2 & 1.716 & 203.2 & 0.68 \\ 
 g2 & 800 & 19176 & 300 & 10000 & 2 & 1.777 & 198.6 & 0.66 \\ 
 g47 & 1000 & 9990 & 300 & 10000 & 2 & 0.337 & 194.3 & 0.65 \\ 
 g23 & 2000 & 19990 & 330 & 23000 & 2 & 1.926 & 666.4 & 2.02 \\ 
 g35 & 2000 & 11778 & 330 & 23000 & 3 & 2.697 & 951.3 & 2.88 \\ 
\bottomrule
\end{tabular}
\end{table}

%% file: tables/gset_best_overall_table.tex
% Auto-generated by render_best_overall_table_edges.py (cuts mode)
\begin{table}[t]\centering
\caption{
Best raw cut values on \textsc{Gset} (higher is better). 
$D^\ast$ denotes the depth selected by best median value across seeds. 
BK: best-known cut from the Breakout Local Search heuristic~\cite{benlic_breakout_2013};
BM-SS: single-shot Burer–Monteiro;
BM-MS: best-at-budget multi-start (10\,min);
GRB: Gurobi 10\,min;
SA(2)-LP$+$Gibbs: $\mathrm{SA}(2)$ linear program solved classically and decoded with the same Gibbs sampler.
}
\label{tab:gset_best_overall_raw_cuts}
\small
\begin{tabular}{lrrrrrrrrrr}\toprule
GSET & $N$ & $|E|$ & BK & BM-MS & BM-SS & Gurobi & SA(2)-LP+Gibbs & Ours & $D^*$/2Q & Approx.~ratio \\
\midrule
 g2 & 800 & 19176 & 11620 & 11620 & 11489 & \textbf{11620} & 11544 & 11580 & 2/42 & 0.997 \\ 
 g14 & 800 & 4694 & 3064 & 3060 & 3004 & 3016 & 3033 & \textbf{3051} & 2/42 & 0.996 \\ 
 g47 & 1000 & 9990 & 6657 & 6657 & 6545 & 6584 & 6591 & \textbf{6622} & 2/42 & 0.995 \\ 
 g23 & 2000 & 19990 & 13344 & 13333 & 13148 & 13178 & 13163 & \textbf{13259} & 2/46 & 0.994 \\ 
 g35 & 2000 & 11778 & 7678 & 7660 & 7517 & 7568 & 7572 & \textbf{7602} & 3/69 & 0.990 \\ 
\bottomrule
\end{tabular}
\end{table}